\documentclass[pra,
reprint,
twocolumn,
amsmath,
amssymb,
aps,
showpacs,
floatfix]{revtex4-2}

\usepackage{natbib}
\usepackage{graphicx}
\usepackage[tight]{subfigure}
\usepackage{bm}
\usepackage{dcolumn}
\usepackage{amssymb}
\usepackage{amsbsy}
\usepackage{times}
\usepackage{color}
\usepackage{float}
\usepackage[colorlinks,bookmarks=false,citecolor=blue,linkcolor=red,urlcolor=blue]{hyperref}

\begin{document}

\title{Theoretical description of atomtronic Josephson junctions in an optical lattice}
\author{Manjari Gupta}
\author{H. R. Krishnamurthy}
\affiliation{Center For Condensed Matter Theory, Department of Physics, Indian Institute of Science, Bangalore 560012, India}
\author{J. K. Freericks}
\affiliation{Department of Physics, Georgetown University, Washington, D.C. 20057, USA}
\date{\today}

\begin{abstract}
Experimental realizations of ``atomtronic" Josephson junctions have recently been created in annular traps in relative rotation with respect to potential barriers that generate the weak links. If these devices are additionally subjected to an optical lattice potential, then they can incorporate strong-coupling Mott physics within the design, which can modify the behavior and can allow for interesting new configurations of barriers and of superfluid flow patterns. We examine theoretically the behavior of a Bose superfluid in an optical lattice in the presence of an annular trap and a barrier across the annular region which acts as a Josephson junction. As the superfluid is rotated, circulating super-currents appear. Beyond a threshold superfluid velocity, phase slips develop, which generate vortices. We use a finite temperature strong-coupling expansion about the mean-field solution of the Bose Hubbard model to calculate various properties of such devices. In addition, we discuss some of the rich behavior that can result when there are Mott regions within the system.
\end{abstract}

\maketitle

\section{Introduction}
``Atomtronics" is a development in the field of ultra-cold atoms, where atomic analogs of various electronic circuits are constructed using trapped atoms \cite{Atomtronics, RevModPhys.94.041001} in the place of electrons. There have been theoretical proposals for various structural units such as diodes and transistors \cite{DandT} using trapped Bose-Einstein condensates~(BEC). There are also experimental advancements demonstrating resistive flow \cite{resistance}, hysteresis \cite{Hysteresis} and atomtronic-qubits \cite{Squbit} in such atomtronic circuits.

Especially interesting are possibilities for atomtronic devices connected with persistent current states, such as Josephson junctions. Persistent currents in superfluids and superconductors, including phenomena such as the onset of dissipation in persistent current states due to phase slips and vortices, have been topics of interest for decades. There have been recent developments in cold atomic gas experiments~\cite{GCampbell2013, GCampbell2011, CRyu,StirredSup,Murray,polo2023perspective,Pezzè2024} which provide new avenues for the study of persistent current states. In these experiments bosonic atoms are cooled into a superfluid BEC state inside a 2D annular trap with a radially directed repulsive barrier across the annulus which acts as a superfluid constriction. Using different trapping techniques either the repulsive barrier \cite{GCampbell2013} or the annular trap \cite{GCampbell2011,CRyu}  can be rotated. In either of  the cases, the experiment simulates superfluid flow through a weak link. The weak link constitutes a Josephson junction and depending on the barrier parameters gives rise to  phase slips beyond a threshold angular momentum of the rotating superfluid, or equivalently, a threshold superfluid velocity. This can be useful for studies in cold atom based SQUIDs (Superconducting Quantum Interference Devices) and in furthering the development of atomtronics. Recently the current-phase relationship in such superfluid weak links has also been measured through interferometry \cite{CurrPhase}.

Inspired by these experiments, there have been a  number of theoretical studies    exploring the dynamics of ultra-cold atomic superfluid flow in annular traps  with such weak-links,  and the current-phase relations at the weak links in different regimes. These studies are mainly based on the Gross Pitaevski equation (GPE), either the time-independent~\cite{Piazza} or the time dependent~\cite{Murray, Piazza2, Piazza3,Gallemi2016, 1DringGasDecay, FewBAnn, asquidTorr,ARIVAZHAGAN20231354180,PhysRevResearch.6.013104,atoms11080109} versions depending on the questions being addressed, and mostly at zero temperature. Most of this work has been done in the context of 1-dimensional or quasi-one dimensional systems \cite{asquidTorr}. Persistent currents in multicomponent Bose gases have also been studied using the GPE \cite{CRPHYS_2023__24_S3_A6_0}. Other recent research includes the study of a rotating periodic lattice \cite{PhysRevA.109.023315}, as well as a study at finite temperatures using the truncated Wigner approximation (TWA) \cite{TWA}, which takes into account thermal and quantum fluctuations beyond the GPE. These studies however, are mainly focused on the weak-coupling limit, which is  reasonable for the experiments that they are focusing on.

Given the enormous progress that has been made building quantum emulators of strongly correlated lattice models such as the Bose  Hubbard model (BHM) \cite{Fisher1989, Freericks1994, HRK2009, Manjari} by trapping ultra-cold bosonic atoms in optical lattices,  it would clearly be interesting to explore persistent currents and superflow past weak links in such systems. Introduction of an optical lattice also provides an opportunity for the exploration of  more complex circuits with Mott phases in between the superfluid phases \cite{Manjari}. To our knowledge, there has been rather limited theoretical or experimental work in this direction. Among the ones we are aware of are a couple of  studies exploring different interaction regimes of a superfluid ring with a constriction using 1D lattice based models \cite{SqubitT,PhasDiag}. In particular, in Ref. \cite{PhasDiag} the phase diagram of the BHM in a 1D ring with tunable local hopping has been studied using quantum Monte Carlo simulations. Other studies of the BHM in ring geometries include the mean-field studies in Ref. \cite{PhysRevB.95.054505} as well the studies in Ref. \cite{PhysRevA.101.023617} using tensor network techniques. There have also been studies of two-component \cite{Gallemi2016} as well as dipolar bosons \cite{PhysRevA.103.013313} without the introduction of a lattice. An interesting related work is the study of persistent currents in the Fermi Hubbard model in a ring shaped geometry at finite temperatures using Bethe Ansatz techniques \cite{PhysRevLett.128.096801}.

This is the main motivation for  the  theoretical study undertaken in this paper, where we consider ultra cold bosons in an optical lattice with an overall annular trap, modelled by an appropriately modified Bose Hubbard Hamiltonian, under conditions of superflow, and in the presence of weak links. We create a weak link using a repulsive barrier potential across the annulus. We focus on the large $U$ limit, where Mott physics can occur, and where the strong-coupling expansion is accurate. If we consider the regime where the average density of atoms is low,  the physics of the lattice model is essentially the same as in the continuum model appropriate to the experiments mentioned earlier.  The strong-coupling expansion about a mean-field treatment of the Bose Hubbard model at finite temperatures~\cite{Manjari} proves to be a good technique for both the low density and the high density regimes, including Mott phenomena, as we describe here. Although the strong coupling techniques we use are powerful enough to address dynamics, we confine ourselves to exploring steady state or ``equilibrium" superflows.

Specifically, we show in this paper that phase slip phenomena can be simulated in optical lattice systems with an overall annular trap potential, both with or without a weak link created by a repulsive ``barrier'' potential, in the strong coupling regime. For the system with a barrier, we produce a ``phase diagram'' for the ``critical current'' for generating phase slips as a function of the barrier potential and the temperature.  We show that phase slips can be seen even in a rotationally symmetric (i.e., no barriers) case but only close to the critical temperature for the superfluid-to-normal transition, occasionally accompanied by vortices.  We also demonstrate that in the presence of high density configurations with Mott phases in between superfluid phases, a variety of superfluid circuits can be generated by the addition of appropriate repulsive (barrier) and attractive (well) potentials, which could be useful in the further development of atomtronics.

The rest of this paper is organized as follows. In the next section we present the BHM with an annular trap potential and a barrier potential on a two dimensional lattice that we use for exploring phase slips and superflow. In Sec. III-A, we discuss the extension of the techniques for strong-coupling perturbation calculations about the mean-field solutions at finite temperatures required to include the presence of persistent currents. In Sec. III-B, we present and discuss the results of our calculations for phase slips, critical currents and vortices. Sec. III-C contains a discussion of the behavior of the critical-current for temperatures near the superfluid-normal transition temperature. In Sec. IV, we discuss some of the possibilities for superfluid flow configurations in circumstances when there are Mott phases in between superfluid phases. Sec. V contains some concluding comments.

\section{Bose Hubbard Model in an annular trap with a barrier}
We use a theoretical model with a lattice based structure to explore persistent currents and phase-slips in the presence of an optical lattice as well as an annular trap and a weak link.  The bosons trapped in the different optical lattice wells are modelled in terms of a BHM on a 2D square lattice, with an additionally imposed overall annular trap potential $V_A({\textbf{r}_j})$, as well as a repulsive barrier potential $V_B({\textbf{r}_j})$ across the annulus which acts as a weak-link, with $\textbf{r}_j$ being the spatial coordinate of the $j$th lattice site. Thus bosonic atoms hop between nearest neighbor sites with amplitude $-t$, interact with an on site repulsion $U$, and are subjected to a local potential $V_A(\textbf{r}_j)+V_B(\textbf{r}_j)$ as well a global chemical potential $\mu$. The full Hamiltonian of our model is therefore

\begin{eqnarray}
\nonumber
\mathcal{H}=&&-\sum_{jj'}t_{jj'}b^{\dagger}_{j}b_{j'}+\frac{U}{2}\sum_j n_{j}(n_{j}-1)-\mu \sum_j n_{j} \\
&&+\sum_j  \left( V_A(\mathbf{r}_j) + V_B(\textbf{r}_j) \right) n_j .
\label{eqH}
\end{eqnarray}
Here $b^{\dagger}_{j}$ ($b_{j}$) is the creation (annihilation) operator for the boson  at site $j$, and $n_{j}=b^{\dagger}_{j} b_{j}$ is the appropriate number operator.

The  annular trap potential $V_{A}$ is modelled by
\begin{equation}
V_A(\mathbf{r}_j)=-V_a \left(\frac{|\mathbf{r}_j|}{r_0}\right)^2 e^{-\left(\frac{|\mathbf{r}_j|}{r_0}\right)^2}.
\label{eqVA}
\end{equation}
The parameters $V_a$ and $r_0$ are varied to achieve different sizes of the annular ring. The repulsive barrier potential $V_B$ across the annular region, which acts as a weak-link, is taken to be a Gaussian function only of $\theta_j$, the angle between $\textbf{r}_j$ and the positive x-axis:
\begin{equation}
V_B(\theta_j)=V_b e^{-\left(\frac{\theta_j}{\theta_0}\right)^2}.
\label{eqVB}
\end{equation}
We will henceforth refer to this potential as the {\em angular}-Gaussian barrier.
Within a local density approximation (LDA), the potentials $V_A(\mathbf{r}_j)$ and $V_B(\theta_j)$ can be thought of as an effective local chemical potential  given by $\mu_j=\mu-V_A(\mathbf{r}_j)+V_B(\theta_j)$.

In the experiments, the barrier is usually created by a laser dot scanned rapidly across the annulus \cite{GCampbell2013,GCampbell2011}. Our angular-Gaussian barrier potential in Eq. (\ref{eqVB}) is a reasonably good choice for modeling the experimental potential provided  the width  $\theta_0$ of the Gaussian is small compared to $2\pi$, whence the spatial width of the barrier potential across the radius of the annulus will be small compared to the average circumference of the annulus. For our low density  calculations (i.e. the ones without Mott regions inside the trap, see below), we have taken $\theta_0=0.1$.

\section{Persistent currents and Phase slips via Strong Coupling Expansion}
Our goal is to theoretically explore persistent current and phase-slip phenomena in the two-dimensional system of bosonic atoms in the presence of an optical lattice, an annular trap and a weak link, modeled by the Hamiltonian in Eq. (\ref{eqH}). We do this by extending, as discussed below, the finite-temperature strong-coupling ($t \ll U$) expansion of the inhomogeneous BHM in the presence of superfluidity discussed in detail in Ref.~\onlinecite{Manjari}. Specifically, we present calculations for  the expectation values of the inhomogeneous, site specific,    density $\langle{n_j\rangle}$, the superfluid order parameter $\langle{b_j\rangle}$, as well as the   local current operators, in states with superflow, up to second order in the perturbation series in $t/U$.

\begin{figure}[t!]
\includegraphics[width=8.5cm]{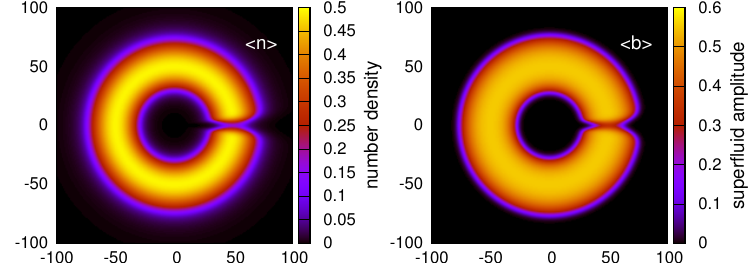}
\caption{(Color online) False color plot of the number density $\langle{n_j\rangle}$ and the superfluid order parameter $\langle{b_j\rangle}$  for the ``non-rotating" case ($l=0$, see text). The left panel shows $\langle{n_j\rangle}$ and the right panel shows $\langle{b_j\rangle}$ for an angular-Gaussian barrier with $V_b=2.3 t$ and $\theta_0=0.1$. (For values of the other parameters see the text.)}
\label{Dn_SFD}
\end{figure}

A simple repetition of  the calculations of Ref.~\onlinecite{Manjari} for the Hamiltonian in Eq. (\ref{eqH}) recovers just the equilibrium state of the Bose superfluid in the annular trap with a barrier, \textit{without superflow}.
Fig. \ref{Dn_SFD} shows a false colour plot of the resulting distribution of the number density $\langle{n_j\rangle}$ and the superfluid order parameter $\langle{b_j\rangle}$ over a $201 \times 201$ sized 2D lattice in the presence of the annular as well as the angular-Gaussian barrier potentials. The parameters are chosen such that the maximum value of the density is $\le 0.5$ and there is no Mott region. The parameters used are $U=20 t$, $V_a=20t$, $r_0=50$, the temperature $T = t$, $V_b=2.3 t$ and $\theta_0=0.1$. (We note that here and in the rest of this paper,  we quote temperature in energy units, and all energy parameters in units of the hopping, $t$.) The chemical potential is taken to be $\mu=-8.2t$ and kept fixed. The annular trap, and the suppression of the density and the order parameter at the barrier, are evident in the figure.

\subsection{The strong-coupling perturbation technique in the presence of persistent currents}
In a superfluid (or a superconductor), the persistent current state is a long lived, metastable, non dissipative, current carrying state of the system which has a circulation that is an integral multiple of $h/m$, where $m$ is the mass of the boson and $h$ is Planck's constant. Persistent current states are stable only in multiply connected regions such as an annular ring or a torus. The bosons in the 2-D annular trap described by the Hamiltonian in Eq. (\ref{eqH}) move inside such a multiply connected region. The superfluid order parameter can in general be complex in such circumstances, and the gradient of the phase of the order parameter determines the superfluid flow.

Thus,  we can treat persistent current states with superfluid flow within the framework of our calculations in Ref.~\onlinecite{Manjari} simply by allowing for a complex order parameter  $\langle{b_j\rangle} \equiv|
\langle{b_j\rangle}|e^{i\theta_{\langle{b_j\rangle}}}$ with a non-uniform phase $\theta_{\langle{b_j\rangle}}$. Given that the order parameter has to be single valued, the total change in $\theta_{\langle{b_j\rangle}}$ around a (lattice) path that winds around the annulus has to be an integral multiple of $2\pi$, and the integer is called the winding number, which we will denote by $l$.
In a continuum approximation, we can write  $\oint_\mathcal{R} \vec{\nabla} \theta_{\langle{b_j\rangle}}\cdot d\vec{l}=2\pi l$, where $d\vec{l}$ is an infinitesimal line element along a contour $\mathcal{R}$ around the annulus at a radial distance $R$ from the central lattice site.

Hence we proceed, as in Ref.~\onlinecite{Manjari}, by rewriting the Hamiltonian in Eq. (\ref{eqH}) as:

\begin{equation}
\mathcal{H}\equiv [\tilde{\mathcal{H}}_0+\sum_{jj'}\phi^*_jt^{-1}_{jj'}\phi_{j'}]
+\tilde{\mathcal{H}}_I ,
\label{eqH_sce}
\end{equation}

\begin{eqnarray}
\nonumber
\tilde{\mathcal{H}}_0 &&=\sum_j\left[ \frac{U}{2}n_j(n_j-1)-
\mu_jn_j-\phi_jb^{\dagger}_j-\phi^*_jb_j\right] \\
&&\equiv\sum_j \tilde{\mathcal{H}}_{0j} .
\label{eqH0}
\end{eqnarray}

\begin{eqnarray}
\nonumber
\tilde{\mathcal{H}}_I &&=-\sum_{jj'}t_{jj'}[b^{\dagger}_j-\langle{b^{\dagger}_j \rangle}_{\tilde{\mathcal{H}}_{0j}}][b_{j'}-\langle{b_{j'} \rangle}_{\tilde{\mathcal{H}}_{0j}}] \\
&&\equiv -\sum_{jj'}t_{jj'}\tilde{b}^{\dagger}_j\tilde{b}_{j'} .
\label{eqHI}
\end{eqnarray}
where $\phi_j\equiv \sum_{j'}t_{jj'}\langle{b_{j'} \rangle}_{\tilde{\mathcal{H}}_{0j}}$ is determined self-consistently from $\tilde{\mathcal{H}}_{0j}$. All the expectation values above  are thermal expectation values with respect to $\tilde{\mathcal{H}}_{0j}$, given by 

\begin{eqnarray}
\nonumber
 \langle{b_j\rangle}_{\tilde{\mathcal{H}}_{0j}} & = & \sum_{\tilde{n}} \rho_{\tilde{n}; j} \langle \tilde{n}; j \vert \, b_j \vert \tilde{n}; j \rangle ; \;  \rho_{\tilde{n}; j} \equiv \frac{1}{\tilde{z}_j} e^{-\beta \tilde{\epsilon}_{\tilde{n}; j}} ; \\  
 \tilde{z}_j  &  \equiv &  \textrm{Tr}  \left( e^{-\beta \tilde{\mathcal{H}}_{0j}} \right)  = \sum_{\tilde n} e^{-\beta \tilde{\epsilon}_{\tilde{n}; j}} ; \; \beta \equiv \frac{1}{k_B T} \, .
\label{eqbjzj}
\end{eqnarray}
Here $\tilde{\epsilon}_{\tilde{n}; j}$ and $\vert \tilde{n}; j \rangle$ are respectively the eigenvalues and eigenvectors of $\tilde{\mathcal{H}}_{0j}$, i.e., $ \tilde{\mathcal{H}}_{0j} \vert \tilde{n}; j \rangle = \tilde{\epsilon}_{\tilde{n}; j} \vert \tilde{n}; j \rangle $, with $\tilde{n}$ being an integer indexing the eigenvectors. $\tilde{z}_j$ is  the partition function of  $\tilde{\mathcal{H}}_{0j}$. The self-consistency calculations are typically done iteratively, starting from an initial choice $\langle{b_j\rangle}_{initial}$ , determining $\phi_j$ and numerically diagonalizing $\tilde{\mathcal{H}}_{0j}$ at every one of the sites within a chosen truncated Boson number basis, recalculating $\langle{b_{j} \rangle}_{\tilde{\mathcal{H}}_{0j}}$, and repeating the process until convergence.  For details see Ref. \onlinecite{Manjari}. We will henceforth denote the converged $\langle{b_{j} \rangle}_{\tilde{\mathcal{H}}_{0j}}$ by $\langle{b_{j} \rangle}_0$.

The key change we make here compared to the calculations in Ref. \onlinecite{Manjari} is that we  bias the self consistent solution in favour of persistent current states by choosing the  phase of $\langle{b_j\rangle}_{initial}$ (denoted  by $\theta_{\langle{b_j\rangle}}^{initial}$) to have a nonzero initial winding number  $l_{initial}$. Typically, we choose $\theta^{initial}_{\langle{b_j\rangle}} = l_{initial} \theta_j$, where $\theta_j$ is the polar angle at site $j$.
From the self-consistent $\phi_j$, we calculate the zeroth order $\langle{b_{j} \rangle}_{0}$. Treating $\tilde{\mathcal{H}}_I$ given in Eq. (\ref{eqHI}) as the perturbation, we calculate $\langle{b_{j} \rangle}$ up-to 2nd order in $\tilde{\mathcal{H}}_I$ \cite{Manjari}. The winding number corresponding to the phase of the final $\langle{b_{j} \rangle}$ (obtained from the mean-field approximation plus the second order corrections), denoted by $l$, is the final, ``equilibrium" winding number. We generally find that when  $l_{initial}$ is nonzero, so is $l$.

To calculate the superfluid current, we calculate the thermal expectation value of the operator representing the particle current between two lattice sites $j$ and $j'$, which is given by,
\begin{equation}
\vec{J}_{jj'}=-i (t_{jj'}b^{\dagger}_jb_{j'}-t_{j'j}b^{\dagger}_{j'}b_j)\vec{e}_{jj'} \equiv J_{jj'}\vec{e}_{jj'}.
\label{eqJ}
\end{equation}
Here $t_{jj'}= t_{j'j}^*$ is the corresponding hopping matrix element. For our calculations, we have used a model with only nearest neighbour hopping given by $t$; $\vec{e}_{jj'}$ is a unit vector along the direction from site $j$ to site $j'$. Within the mean-field calculations, the expectation value of the current from site $j$ to $j'$ in the mean-field superfluid state is  given by $\langle{J_{jj'}\rangle}^{(0)}\vec{e}_{jj'}$ where
\begin{equation}
\langle{J_{jj'}\rangle}^{(0)}=-2t|\langle{b_j\rangle}_0||\langle{b_{j'}\rangle}_0|
\sin(\theta_{\langle{b_j\rangle}_0}-\theta_{\langle{b_{j'}\rangle}_0}),
\label{eqJ0}
\end{equation}
showing the well known sinusoidal dependence of the super-current on the variation with position of the phase of the order parameter. For the cases where $\theta_{\langle{b\rangle}}$ varies slowly spatially, the supercurrent can be approximated as being proportional to the gradient of the $\theta_{\langle{b\rangle}}$ field, leading to the standard result that the supercurrent is proportional to the superfluid velocity ($\vec{v_s} \propto \nabla \theta_{\langle{b\rangle}}$).

One can go beyond the mean-field approximation by substituting  $b_j = \langle{b_j\rangle}_0 + \tilde{b}_{j} $ in the expression for the current operator, expand the latter out, and evaluate the thermal expectation values of all the resulting terms (including the additional terms involving $\tilde{b}_{j}$ and $\tilde{b}_j^{\dagger}$) in the thermal ensemble corresponding to the full Hamiltonian $\mathcal{H}$ (cf., Eq. (\ref{eqH_sce})) as a perturbative expansion in powers of $\tilde{\mathcal{H}}_I$ (cf., Eq. (\ref{eqHI})), ideally ensuring that the prescription does not break gauge-invariance and current conservation. The zeroth order term in this expansion is exactly the mean-field  current density in Eq. (\ref{eqJ0}), and we show explicitly in the Appendix that this by itself obeys both gauge-invariance and current conservation. In this paper, we restrict ourselves to presenting results only for this mean-field current density,  as we find that the next (first) order contribution to the current density, while computationally more demanding as regards their calculation, are numerically small ($ < 0.05 \%$ of the zeroth order contributions), and make little qualitative difference to the results in all the contexts we discuss in this paper.

In order to make our calculations able to describe the continuum limit appropriate to the experiments discussed in the introduction, we have chosen parameters so that the average density $\langle{n\rangle}$ is low, and the maximum among the site specific  densities, $\{\langle{n_j\rangle}\}$, is never more than $0.5$, and generally significantly smaller. The number of lattice sites present inside the annular region is made sufficiently large (typically $16,000$)  to avoid finite size effects.   As we change the rotation or the winding number and the height of the barrier potential $V_b$ in our calculations, the total number of particles roughly varies from $4000$ to $6000$ and the entropy per particle is typically $0.5k_B$ to $0.7k_B$.


\begin{figure}[t!]
\includegraphics[width=8.5cm]{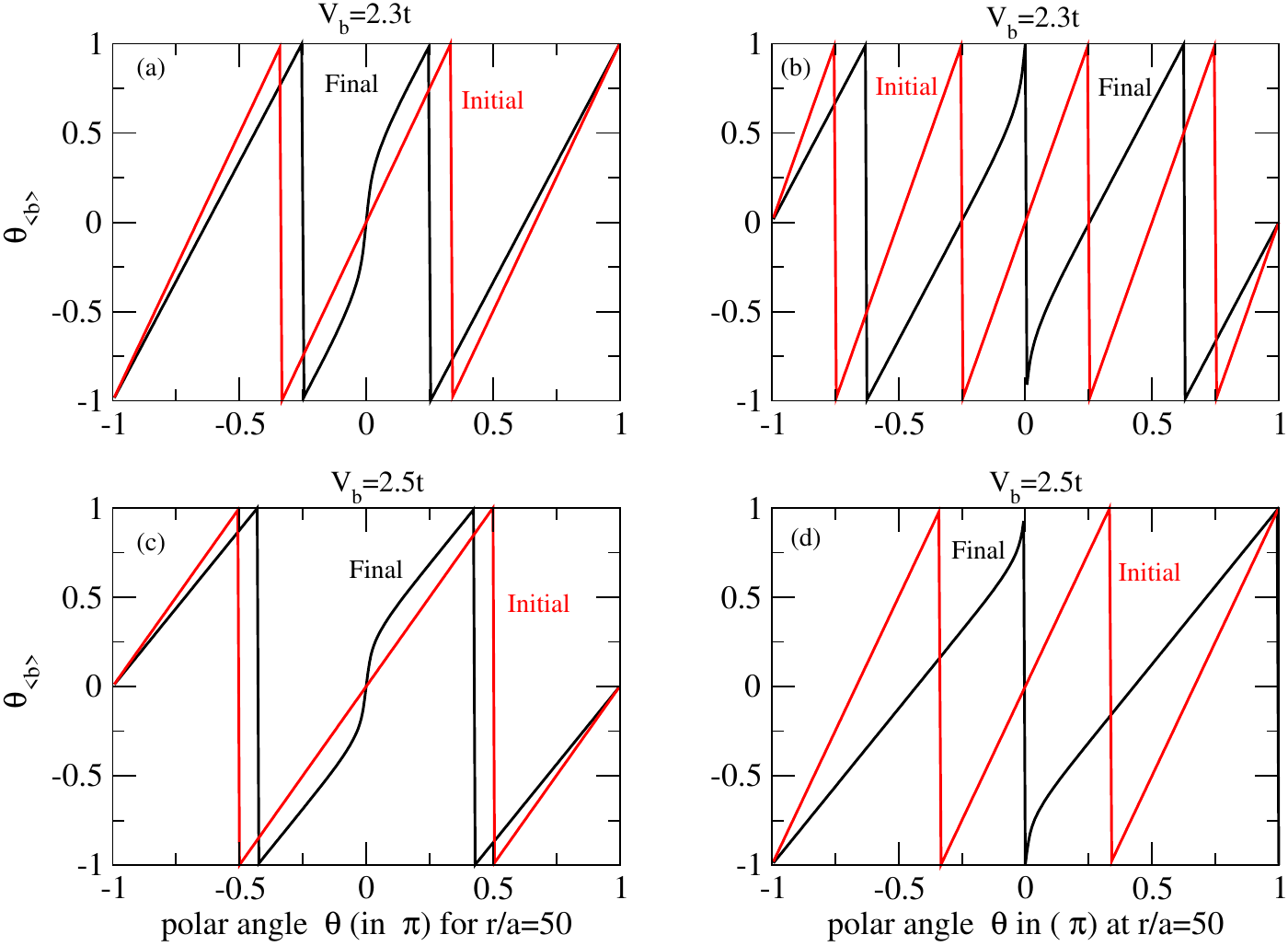}
\caption{(Color online) Plots of $\theta_{\langle{b_j\rangle}}$, the argument of $\langle{b_j\rangle}$,  as functions of the azimuthal angle of the sites around the annulus at a radial distance $r/a=50$. The upper panel shows $\theta_{\langle{b\rangle}}$ before ($a$) and after ($b$) phase slip ($l_{max}=3$) for $V_b=2.3t$ and the lower panel shows $\theta_{\langle{b\rangle}}$ before ($c$) and after ($d$) phase slip ($l_{max}=2$) for $V_b=2.5t$. $\theta_{\langle{b\rangle}}$is  in units of $\pi$ and has been constrained to be in the range from $-\pi$ to $\pi$.}
\label{PhaseSlip}
\end{figure}

\subsection{Critical currents, Phase slips and Vortices}
It is well known \cite{QSDnDecay} that current carrying states of super-fluids are (metastable) states of higher free energy than the equilibrium state, and when the superfluid current (or velocity) exceeds a critical value the system goes normal (even in a homogeneous system - see subsection C) .  In the present case of steady-state currents in an annular system with a barrier, clearly the supercurrent is proportional to the self-consistent phase winding around  the annulus, i.e., to $l$.  Typically, if $l_{initial}$ is small enough, we find that  $l=l_{initial}$, and hence it can be increased by increasing $l_{initial}$. This procedure is the theoretical analog of increasing the relative angular velocity between the annulus and the repulsive barrier in the experimental set-up. In the system with a barrier the value of the critical current is known to decrease with increasing barrier height. Generally, when the initial phase winding is such that the corresponding supercurrent is lower than the critical current, the final or the self-consistent phase winding is the same as the initial winding. If the initial phase winding gives rise to a superfluid velocity which is greater than the critical velocity near the barrier region, the self-consistent phase winding drops by integral multiples of $2\pi$, leading to the phenomenon of phase slip. Thus, as $l_{initial}$ is increased beyond a threshold $l_{max}$ which depends on the system parameters,  $l$ drops below $l_{initial}$ and sticks at $l_{max}$, due to phase slips.

Fig. \ref{PhaseSlip} shows examples of  this phase slip phenomenon for two different barrier heights $V_b=2.3t$ (shown in the upper panel) and $V_b=2.5t$ (shown in the lower panel). We find that for $V_b=2.3t$ and $V_b=2.5t$ the maximum superflow possible, i.e., the critical current, is reached for the winding numbers $l_{max}=3$ and $l_{max}=2$ respectively. If the initial winding number $l_{initial}$ is greater than these values of $l_{max}$, then phase slips set in. Note that even before the occurrence of a phase slip, the phase of the order parameter changes more rapidly in the barrier region. This is necessary because in the steady persistent current state the current across any section of the annulus has to be the same, and the increased phase gradient (or velocity) in the barrier region compensates for the suppression of the superfluid order parameter  there  to ensure current conservation.  Note also that for the same reason the bulk of the phase slip occurs near the barrier region (where $\theta_j \approx 0$). The calculations for the data in Fig. \ref{PhaseSlip} have been  done at a fixed temperature $T= 1t$, with the  chemical potential $\mu$ in Eq. \ref{eqH}, kept fixed at $-8.2t$. The total number of particles then varies from $4000$ to $6000$.

\begin{figure}[b!]
\includegraphics[width=8cm]{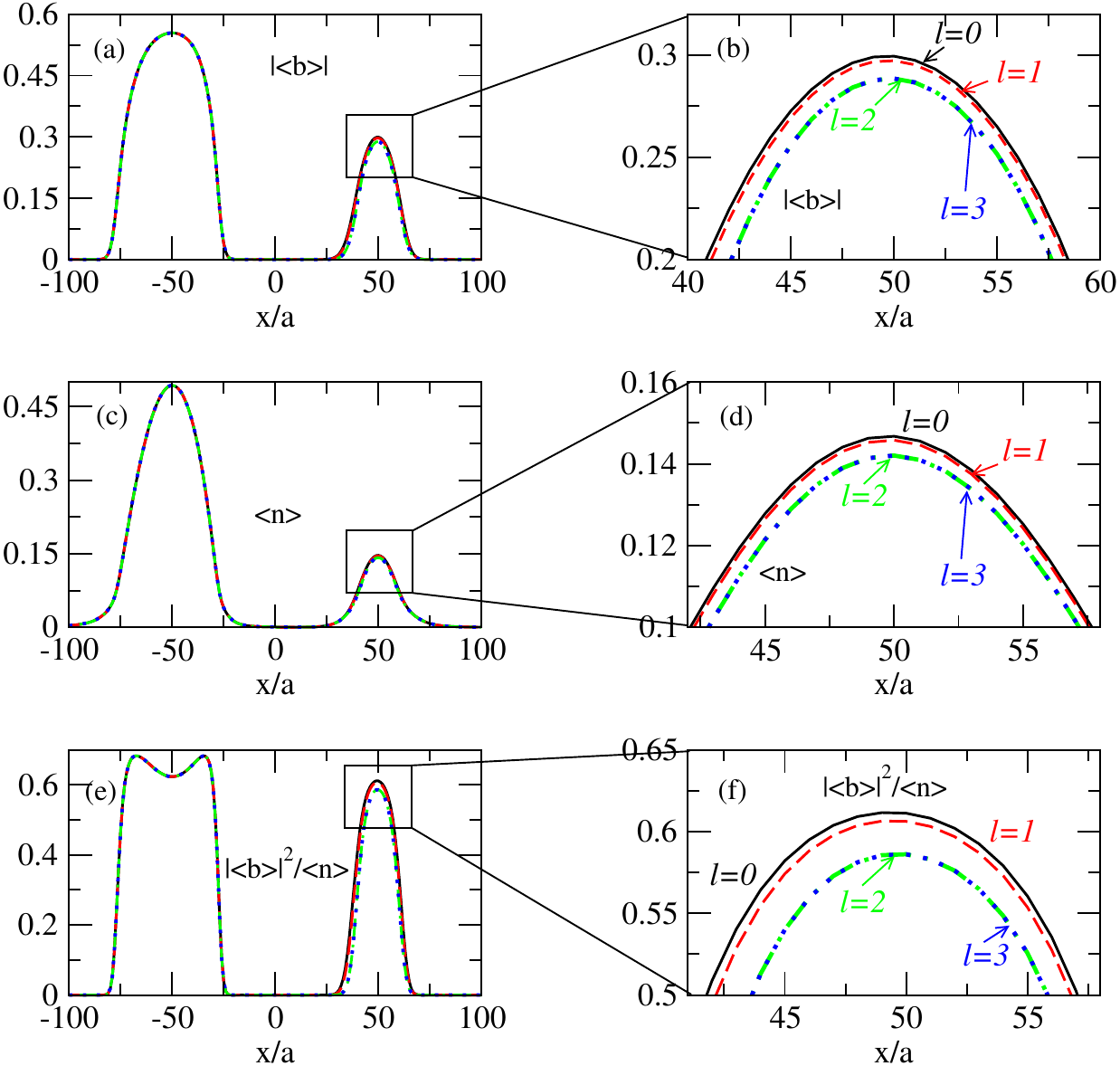}
\caption{(Color online) The number density $\langle{n_j\rangle}$,  the magnitude of the superfluid order parameter $|\langle{b_j\rangle}|$ and the the superfluid fraction  $\rho_{fj}$ plotted along the {\it x}-axis of the annulus for $V_b = 2.5t$ and for different values of the winding number $l$. Fig 3$(a)$ shows $|\langle{b\rangle}_j|$ for $l=0,1,2$. Fig. 3$(b)$ shows a close-up of $|\langle{b_j\rangle}|$ near the central region of the barrier. Fig. 3$(c)$ shows  $\langle{n_j\rangle}$ along the {\it x}-axis for $l=0,1,2$. Fig. 3$(e)$ shows  $\rho_{fj}$ for the different $l$. Fig. 3$(d)$ and Fig. 3$(f)$ shows the close-up versions of $\langle{n_j\rangle}$ and $\rho_{fj}$ respectively along the {\it x}-axis for $l=0,1,2$.}
\label{Dn_SFD_2}
\end{figure}

Fig. \ref{Dn_SFD_2} shows the variation of the absolute value of the superfluid order parameter $|\langle{b_j\rangle}|$, the density profile $\langle{n_j\rangle}$, and the (mean-field) superfluid fraction $\rho_{fj}~\equiv|\langle{b_j\rangle}|^2/\langle{n_j\rangle}$ along the {\it x}-axis of the annulus (the barrier is situated in the positive half of the {\it x}-axis) with changing winding number $l$. We note that as $l$ increases, the value of $|\langle{b\rangle}|$ in the barrier region decreases. This adjustment in the order parameter and the change in the phase gradient are constrained  by the conservation of the super current as mentioned earlier. The density profile $\langle{n\rangle}$ and the superfluid fraction  also  exhibit some dependence on the winding number. For the calculations shown in Fig. \ref{Dn_SFD_2}, the barrier height $V_b=2.5t,$ and the total number of particles is kept fixed at $N=5700$. For these parameter values, the threshold winding number $l_{max}$ is 2, and for winding numbers larger  than $l_{max}$, the profiles for both $|\langle{b\rangle}|$ and $\langle{n\rangle}$ pretty much coincide with their $l=l_{max}$ profiles.

\begin{figure}[t!]
\includegraphics[width=7cm]{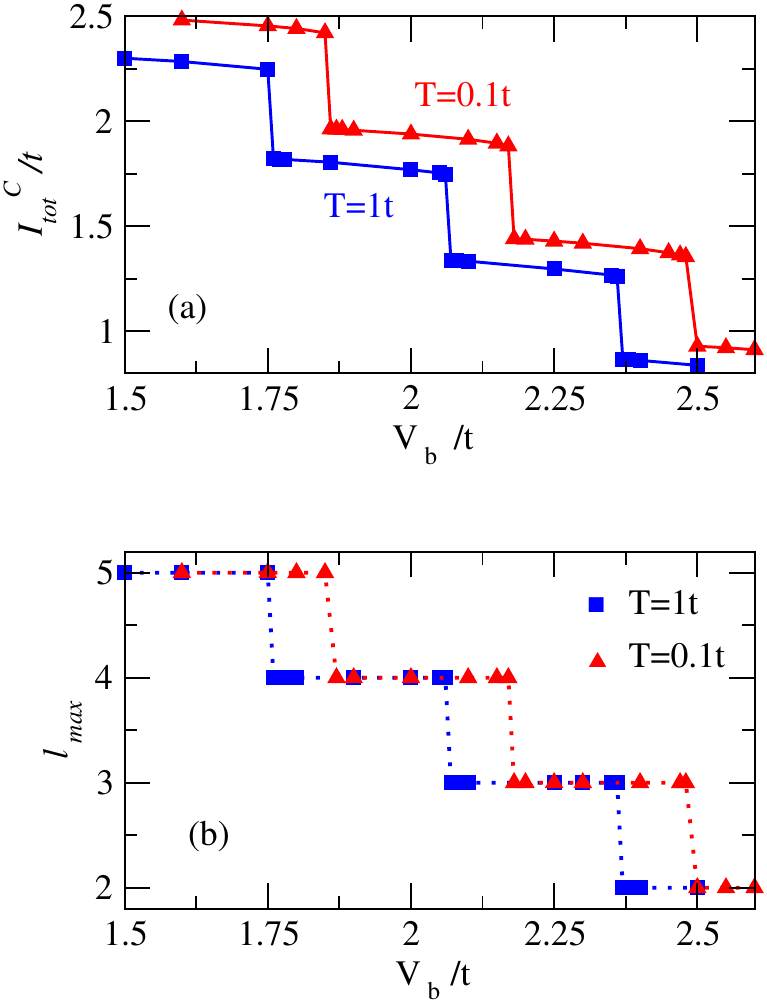}
\caption{(Color online) Fig. \ref{nVb}$(a)$: critical values of the total current across the width of the annulus ($I^c_{tot}$, in units of $t$) plotted against $V_b$ (in units of $t$) for two different temperatures, $T=1t$ and $T=0.1t$. Fig. \ref{nVb}$(b)$: maximum possible winding number ($l_{max}$) beyond which phase slip occurs as a function of $V_b$, for the same two temperatures.  }
\label{nVb}
\end{figure}

The critical current, as well as the threshold winding number, $l_{max}$, for the onset of phase slip depend   on the barrier height $V_b$ as well as on the temperature ($T$) of the system. This dependence is exhibited in Fig. \ref{nVb}.  Fig \ref{nVb}$(a)$ shows the critical value of the total current  circulating around  the annulus (obtained by adding up the expectation values of the current across all the bonds  cut by a radial line from the center of the annulus), $I^c_{tot}$, as a function of the barrier height $V_b$ for two different temperatures $T = 1t$ and $T=0.1t$.   In Fig. \ref{nVb}$(b)$, we plot  $l_{max}$ versus the barrier height $V_b$ for the same two temperatures $T = 1t$ and $T = 0.1t$. As is clear from these figures, both $I^c_{tot}$ and $l_{max}$ generally decrease with increasing temperature at a fixed barrier height, or with increasing barrier height for a fixed temperature, although, in case of the latter, because it takes only integer values, the change shows up as overlapping steps. Thinking of persistent current states with different values of $l$ as different ``phases'', the above pictures can be viewed as depictions of ``first order" phase transitions between these states, occurring when their free energies cross.

\begin{figure*}[th!]
\includegraphics[height=4.2cm,width=5.5cm]{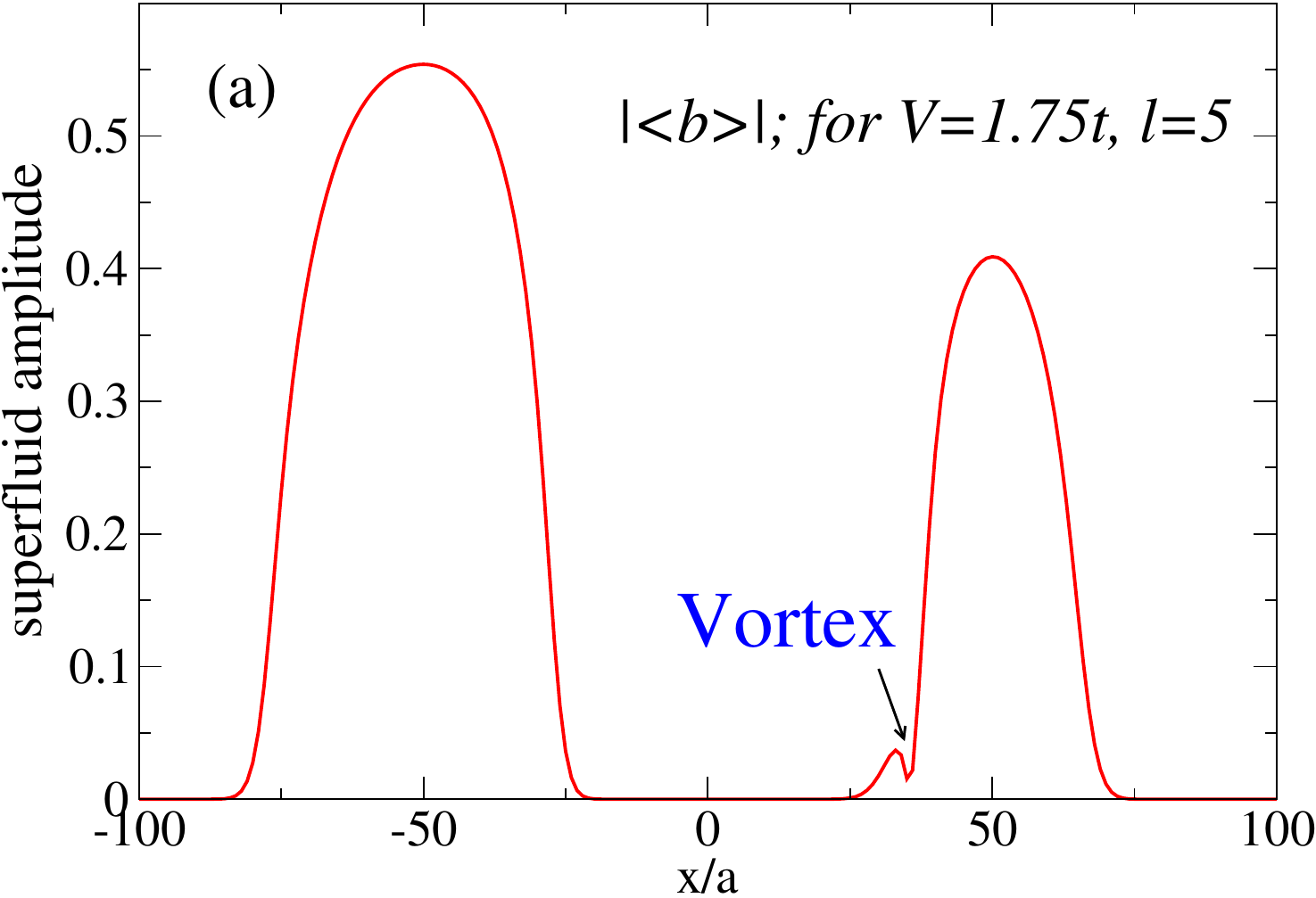}
\hspace{1cm}
\includegraphics[height=5cm,width=7cm]{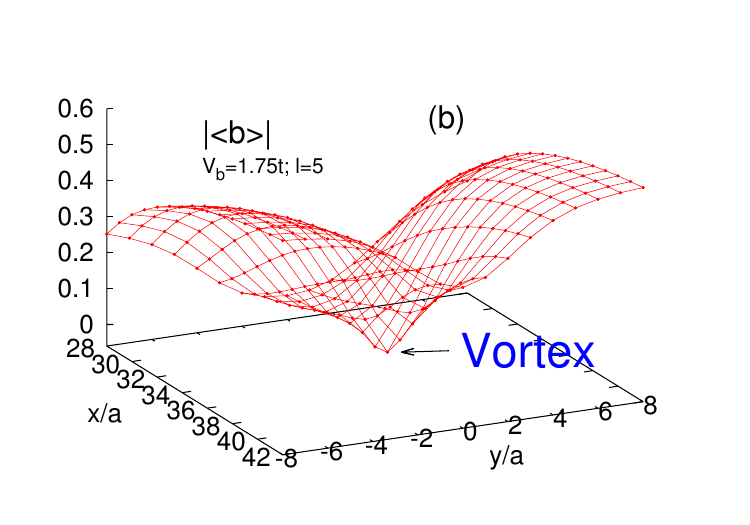} \\
\includegraphics[height=5cm,width=7cm]{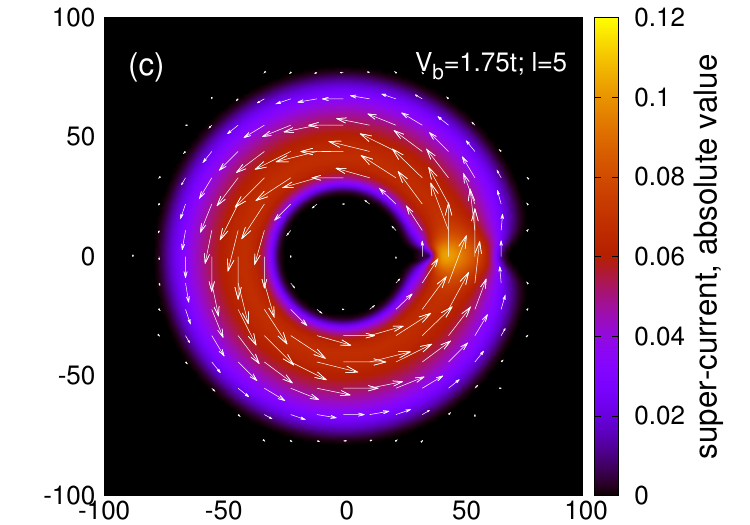}
\hspace{1cm}
\includegraphics[height=5cm,width=7cm]{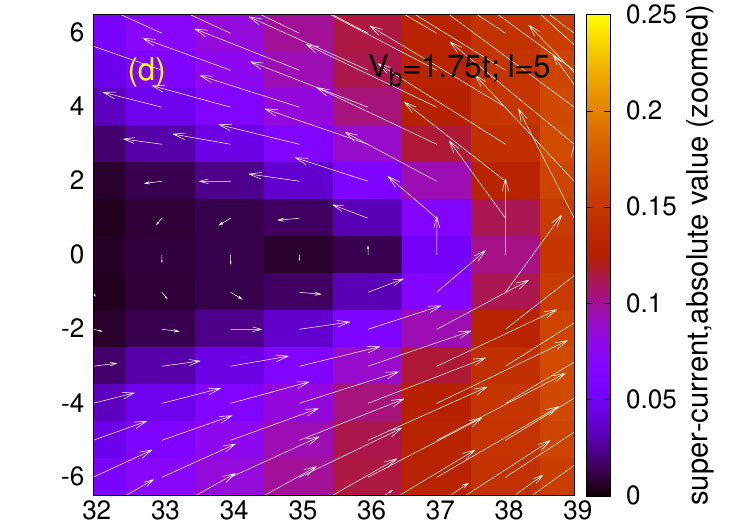}
\caption{(Color online) The absolute value of the order parameter, $|\langle{b_j\rangle}|$, shown in  \ref{Vortex}$(a)$, shows a dip near the inner edge of the annulus at the barrier region near $x=35,y=0$. The profile of this dip (in $|\langle{b_j\rangle}|$) is more clearly shown in a 3D plot in  \ref{Vortex}$(b)$. The lower panel shows the superfluid current flow field at a coarse-grained level (\ref{Vortex}$(c)$) and without coarse-graining (\ref{Vortex}$(d)$), the latter clearly showing the vortex in the current pattern centered around $x=35,y=0$. The arrows show the direction of the current vectors at the appropriate level of coarse-graining, and their lengths are representative of  the relative magnitudes of the current vectors. The false colors in the lower panel represent the absolute value for the net current flow without coarse-graining, i.e. for a square region containing a single lattice site.}
\label{Vortex}
\end{figure*}

Studies of the dynamics of phase slips in a weak coupling BEC system  using the time dependent GPE  mentioned in the introduction (\cite{Piazza2,Piazza3}), have shown  that the generation and motion of vortices and anti-vortices are crucially responsible for phase slips. Typically, for specific combinations of the barrier height, temperature, and super flow, i.e., when the conditions are ripe for a phase slip to occur, a vortex forms at the inner edge of the annulus accompanied by an anti-vortex at the outer edge of the annulus. The crossing of one vortex or anti-vortex across the annulus gives rise to the observed phase slip. Although the  strong-coupling expansion can be and has been extended to include time dependent and non-equilibrium processes, the calculations presented here are essentially in ``equilibrium", and do not include the time-dependent mechanisms of dissipation leading to the generation of vortices at the onset of a phase slip. An interesting question nevertheless is whether vortex like structures ever arise in our calculations.

In order to explore this possibility, we visualize the current configurations and flow patterns arising in our calculations  by constructing  coarse-grained versions of these microscopic currents at various length scales as follows. We divide our 2D square lattice into square blocks with an odd number of sites at each side. The boundary of each block cuts through the bonds. We calculate the net (incoming or outgoing) current flow through the block and assign that vector to the central site of the block. We then pictorially represent the resulting vector field, scaling the lengths of the vectors shown in the figure to be proportional to their magnitudes, i.e., of the (coarse grained) currents. The minimum block size consists of one lattice site which corresponds to no-coarse-graining.

Using such visualization, we have found self-consistent solutions with vortex like metastable structures in some of our calculations when the currents are close to the critical current, and in regions where the value of $|\langle{b\rangle}|$ is sufficiently low. Specifically, in our calculations, we have seen these structures only at the inner edge of the annulus in the barrier region, and have not found an instance where it is accompanied by an anti-vortex at the outer edge of the annulus as found in the time dependent GPE studies mentioned above. Fig \ref{Vortex} shows the various features of one such  solution with a vortex like structure, at the inner edge of the barrier region, for $V_b=1.75t$ and $\theta_0=0.1$. The upper panel shows the dip in the order parameter at the {\it x}-axis near $x/a=35$. The lower panel shows the current patterns. The coarse grained currents in  Fig. \ref{Vortex}$(c)$ depict the  circulating supercurrents, but one needs  the non-coarse grained, microscopic current pattern, shown in  Fig. \ref{Vortex}$(d)$, to be able to clearly make out the vortex.


\begin{figure}[t!]
\includegraphics[width=8.5cm]{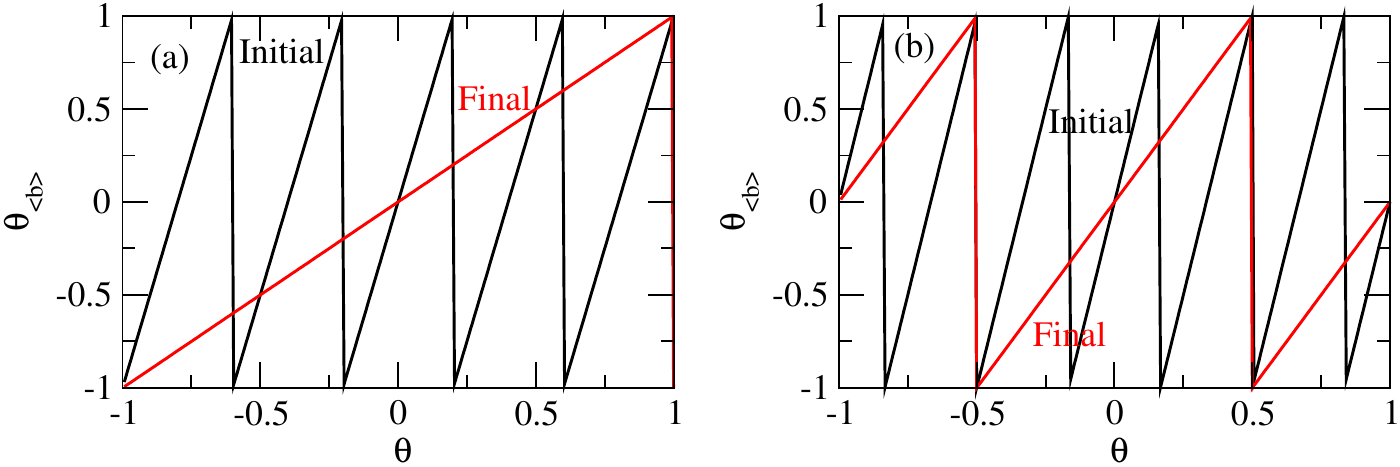} \\
\includegraphics[width=8.8cm]{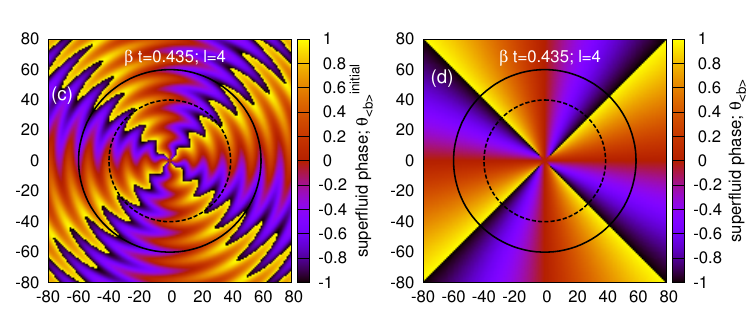} \\
\includegraphics[width=8.5cm]{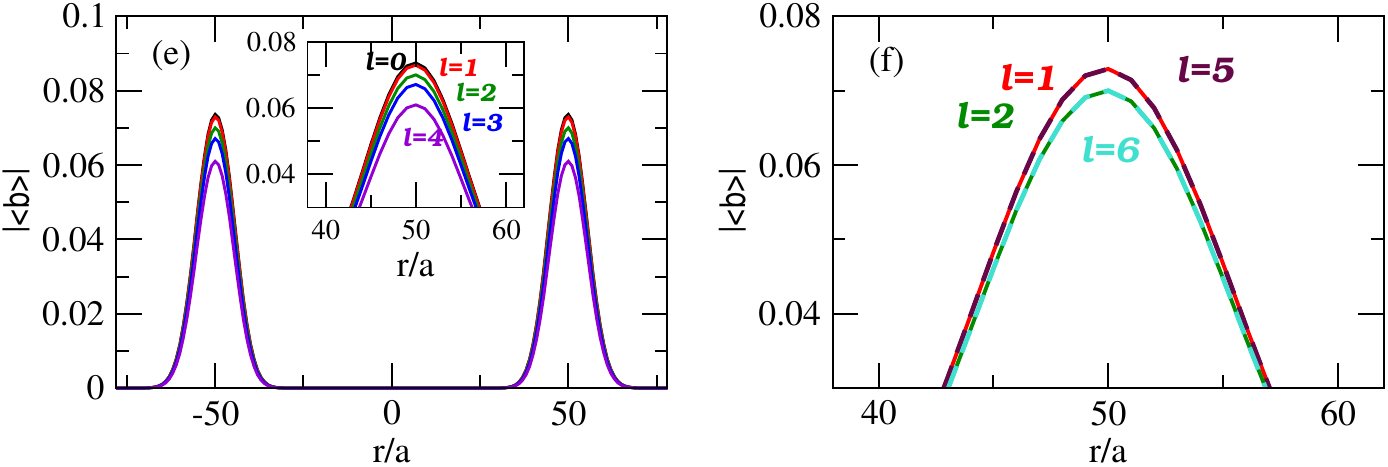}
\caption{(Color online) Results for the annular trap in the absence of the barrier for $\beta=0.435/t$, corresponding to a temperature close to but below the critical temperature.  Panels$(a)$ and $(b)$ show the initial and the self-consistent phase winding around the annulus for two different values of  $l_{initial}$ greater than $l_{max}$. Panels $(c)$ and $(d)$ show, in a false color plot, the initial and self-consistent phases on the lattice for $l=l_{max}=4$.  $(e)$: the superfluid order parameter ($|\langle{b\rangle}|_j$) along the $x$ axis for different winding numbers. $(f)$: the order parameter profile for initial winding numbers ($l_{initial}$) greater than the critical winding number $l_{max}=4$.\\ }
\label{NearTc}
\end{figure}

\subsection{Behavior near the critical temperature and critical super-current}
As is  clear from Fig. \ref{Dn_SFD_2}, the superfluid density decreases as the velocity of the superfow increases, but primarily only in the barrier region. Outside the barrier region ($V_B$),  the reduction in the superfluid density with the increase in the phase winding $l$ is negligible. This is closely connected with the fact that it is difficult to generate a phase slip for the rotationally symmetric case (i.e. without the barrier)   unless the calculations are done close to the critical temperature; otherwise, the corresponding critical phase winding number $l_{max}$ is much too large, comparable or greater than the number of lattice sites around the annulus at a certain radius for all of the annular region.

It is interesting to explore (using strong-coupling mean-field theory) the behavior of the density profile and the order parameter near the critical temperature as the winding number is increased in an annular trap in the absence of the barrier. Fig. \ref{NearTc} shows some of our results, obtained for $\beta=0.435/t$, which corresponds to a temperature close to but lower than the critical temperature for the chemical potential $\mu=-8.2t$. The critical superflow for $\beta=0.435/t$ for the same trap parameters as used earlier occurs at $l_{max}=4$. When we increase the initial phase winding number beyond $l_{max}$ the self-consistent phase winding slips by integral multiples of $2\pi$ and the final winding number becomes less than $l_{max}$, as depicted in Figs. \ref{NearTc} $(a)$ and $(b)$, with no vortices showing up in the final ``equilibrium" configuration. In Figs. \ref{NearTc}$(a)$ and $(b)$, the initial ($l_{initial}$) and the self-consistent ($l$) winding numbers, written in the form ($l_{initial}$,~$l$) are ($5$,~$1$) and ($6$,~$2$) respectively. For the rotationally symmetric case (i.e. without the barrier), the phase of the self-consistent order parameter $\theta_{\langle{b_j\rangle}}$ is simply $l \theta_j$   everywhere on the lattice, as is to be expected from symmetry considerations. The same final configurations arise even if the initial phase winding is not chosen to be $l_{initial} \theta_j$.    Figs. \ref{NearTc}$(c)$ and $(d)$ depict, in the form of false color plots,  the initial and self-consistent phase of the order parameter on the lattice for $l=l_{max}=4$, when the initial value of the phase is chosen to have some jitter as a function of the radius of the lattice point, eg., 

\begin{equation}
\theta^{initial}_{\langle{b_j\rangle}}=l_{initial}[|\frac{sin(\frac{|\vec{r}_j|}{5.0})}{5.0}|+\theta_j], 
\end{equation}
where $j$ is the lattice site index. The annular region, shown by the dotted lines, shows the area where the superfluid order parameter is non-zero for $\beta=0.435/t$. In Figs. \ref{NearTc} $(e)$ and $(f)$, we show the changes in the absolute value of the superfluid order parameter (along the x axis, equivalently any axis through the center of the annulus) with the increase in winding number around the annulus.  In Fig. \ref{NearTc}$(e)$, we show that $|\langle b_j \rangle|$ decreases monotonically throughout the annulus (the inset shows a close up of  the  $|\langle b_j \rangle|$ profile along the positive {\it x}-axis), until the critical current for phase slippage is reached at $l=4$. Within our theory, when $l_{initial}$ is greater than $l_{max}$ the absolute value of the superfluid order parameter does not decrease any further or go to zero; instead for $l_{initial} > l_{max}$, when the self-consistent winding $l < l_{max}$, the self-consistent absolute value of the superfluid order parameter $|\langle{b\rangle}|$ also coincides with the value for $l_{initial}=l$, when $l_{initial} = l < l_{max}$, as shown in Fig. \ref{NearTc} (b).


\begin{figure*}[t!]
\includegraphics[width=18cm]{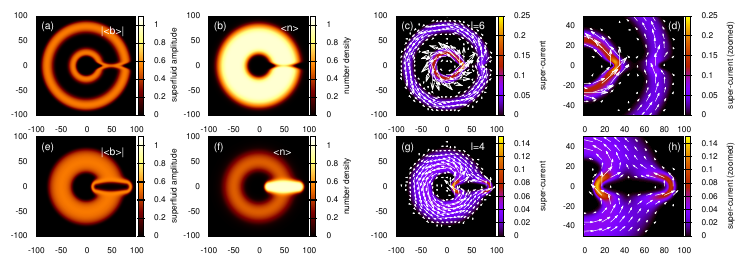}
\caption{(Color online) Two different configurations which include a single $\langle{n\rangle} = 1$ Mott region are shown - one with a concentric Mott region around the annulus (upper panels:(a), (b), (c) and (d)), and the other with the Mott region at a \textit{trench} (lower panels:(e), (f), (g) and  (h)). ((a), upper; (e), lower): false color plots of the order parameter $\langle{b_j\rangle}$; ((b), upper; (f), lower): false color plots of the number density $\langle{n_j\rangle}.$ The right panels, in both the cases, shows the coarse-grained supercurrents in configurations with superflow. For the upper panels, $V_a=50t$, $V_b=6t$, and $U=30t$; for the lower panels, $V_a=20t$, $V_b=-13t$ and $U=25t$. The barrier potential (upper panels) is of the form in Eq. (\ref{rect-gauss-pot}), with width  $d_0=5$. The trench potential (lower panels) is of the form in Eqs. (\ref{rect-gauss-pot}) \textit{and} (\ref{trench-addon-pot}), with $d_0=10$.}
\label{Mott1}
\end{figure*}

\subsection{Comparison with the Gross Pitaevski Equation \\ approach }
The mean field approximation for the Bose Hubbard model is closely related to a discretized version of the Gross Pitaevski equation,  with the discretization parameter for the GPE being equal to the lattice spacing in the  BHM. We discuss this connection below.

A square lattice discretized version of the GPE for Bosons of mass $m$ in two spatial dimensions, with a discretization parameter $a$, can be written as:
\begin{eqnarray}
\nonumber
i\hbar \frac{\partial \Psi_{i,j}}{\partial t}&&=-\frac{\hbar^2}{2ma^2}
[\Psi_{i+1,j}+\Psi_{i-1,j} \\
\nonumber
&&+\Psi_{i,j+1}+\Psi_{i,j-1}-4\Psi_{i,j}] \\
&&+V_{i,j}^{ext}\Psi_{i,j}+g\Psi^{\dagger}_{i,j}\Psi_{i,j}\Psi_{i,j}
\label{eq_SL_GPE}
\end{eqnarray}
where, for clarity, we have denoted the sites of the square lattice by a double rather than a single index. Here $g$ is the strength of the interaction between the bosons modeled as a contact interaction, and is given by $g=\frac{4\pi\hbar^2a_s}{m}$ where $a_s$ is the s-wave scattering length, and $V_{i,j}^{ext}$ is the external potential the Bosons are subjected to.

The Heisenberg equation of motions  for the destruction operator of the BHM given by Eq. (\ref{eqH}), in a similar notation, is given by:
\begin{eqnarray}
i\hbar \frac{\partial b_{i,j}}{\partial t}&&=-t[b_{i+1,j}+b_{i-1,j}+b_{i,j+1} \nonumber \\
&&+b_{i,j-1}]-\mu b_{i,j}+ V_{i,j} b_{i,j}+Un_{i,j}b_{i,j}
\label{bhm-eom}
\end{eqnarray}
where, now, $V_{i,j} \equiv V_A(\vec{r}_{i,j}) + V_B(\vec{r}_{i,j})$.
If we take the expectation value of Eq. (\ref{bhm-eom}), and replace the expectation value of the last term by the product of the expectation values, the two equations become identical, with the following correspondence:
\begin{equation}
\langle b_{i,j} \rangle \leftrightarrow \Psi_{i,j} ~; t \leftrightarrow \frac{\hbar^2}{2ma^2} ~; U \leftrightarrow g ~;
\end{equation}
and
\begin{equation}
V_{i,j}-\mu-4t \leftrightarrow V_{i,j}^{ext}
\end{equation}


Our mean-field calculations in this paper are in equilibrium, which corresponds to the time independent GPE.  In the limit of weak coupling and low density,  the expectation value of the interaction term can indeed be approximated as the product of the expectation values, and we recover the GPE.
However, contrary to the GPE approach, our mean-field calculations treat the on-site interaction exactly, which goes beyond weak-coupling physics, and in particular, allow us to explore the rich possibilities of superflow in the presence of Mott physics, which we discuss in the next section.


\section{Configurations with Mott phases}
One of the main motivations of this work is to explore persistent current phenomena in annular traps in the presence of  Mott phases. The  strong coupling expansion technique does an excellent job of capturing Mott physics in the BHM. To generate Mott regions which have integer site occupancies (primarily $\langle{n\rangle}=1$ regions) within our model, one needs to increase the filling or the Boson number density. This can be done by a combination of increasing the on-site repulsion $U$ for the Bosons and the depth of the annular potential $V_a$, or by introducing an \textit{attractive trench} instead of a repulsive barrier.   All these configurations provide different possibilities of superfluid circuits  depending on the various  parameters, and we discuss some of them below.

In an annular trap without a barrier, the first scenario (increasing $U$ and $V_a$) leads to concentric Mott and superfluid regions and hence to an enlargement of the width of the annular trap. In this case, calculations with an angular-Gaussian  barrier potential do not lead to a realistic representation of the experiments  as the angular spread of the barrier region can increase substantially  across the width of the annulus.  A more realistic solution (from theoretical calculations) is obtained by the use of a {\em rectangular}-Gaussian barrier, i.e., a barrier potential given by
\begin{eqnarray}
	V_{B}(x_j,y_j)&=&\begin{cases}
		V_b e^{-\left(\frac{y_j}{d_0}\right)^2} & \text{if $x_{max} \ge x_j \ge x_{min}$}\\
		0 & \text{otherwise}
	       \end{cases}
\label{rect-gauss-pot}
\end{eqnarray}
where $x_j,y_j$ are the co-ordinates of the $j$th lattice site.

The second scenario, where the potential $V_B$ is made attractive and converted into a trench, needs additional modifications apart from $V_b$ just being made negative in a potential of the form given by Eq. (\ref{rect-gauss-pot}). Outside the edges of the trench region at $x_{min}$ and $x_{max}$, ``semi-circular'' 2D Gaussian potentials of the same width need to be added, so that the trench potential goes to zero gradually rather than suddenly:
\begin{eqnarray}
	V_{B}(x_j,y_j)&=&\begin{cases}
		V_b e^{-\left(\frac{x_j-x_{min}}{d_0}\right)^2-\left(\frac{y_j}{d_0}\right)^2} & \text{if $x_j \le x_{min}$}  \\
		V_b e^{-\left(\frac{x_j-x_{max}}{d_0}\right)^2-\left(\frac{y_j}{d_0}\right)^2} & \text{if $x_j \ge x_{max}$} \\
	  0 & \text{otherwise}
             \end{cases}
\label{trench-addon-pot}
\end{eqnarray}

We have also explored superfluid circuit configurations that arise from the inclusion of  a \textit{second}  rectangular-Gaussian barrier perpendicularly cutting the  barrier modelled by the potential of Eq. \ref{rect-gauss-pot} at a generic point $x_0$ along the positive $x$-axis, i.e., of the form:
\begin{eqnarray}
	V_{B}(x_j,y_j)&=&\begin{cases}
		V_b e^{-\left(\frac{x_j-x_0}{d_0}\right)^2} & \text{if $y_{max} \ge y_j \ge y_{min}$}  \\
		0 & \text{otherwise}
	        \end{cases}
\label{perp-barr-pot}
\end{eqnarray}

\subsection{Configurations with a single $\langle{n\rangle}=1$ Mott  phase}
As mentioned above, there are two different scenarios by which Mott regions  can be generated  inside the annular trap. In the first scenario,  the depth of the annular trap $V_a$  as well as the on site repulsion  $U$ are increased. In this case, without the barrier one gets concentric and alternating rings of Mott and superfluid regions.  The simplest configuration is having one Mott ring with $\langle{n\rangle}=1$ in between two superfluid rings. Addition of a repulsive barrier can give rise to different situations depending on the strength of $V_b$. If  $V_b$ is large  the two superfluid rings get connected by the  generation of superfluid channels along the barrier, or if $V_b$ is small the two superfuid rings get slightly distorted but remain disconnected.

In  the second scenario, if $V_B$ is made attractive, converting what was a barrier into a trench, a patch of the Mott phase with $\langle{n\rangle}=1$ is created at the trench region, surrounded by the superfluid regions. In this case $U$ need not be increased as much as for the previous case and the overall trap potential $V_a$ can  also be kept fixed. The filling remains much less than $1$ around the annulus, except in the trench region.

Fig.~\ref{Mott1} depicts the results from our calculations for these two different scenarios. For the first scenario (upper panels) the parameters are $V_a=50t$,  $U=30t$, $V_b=6t$, $d_0=5$, and the chemical potential is fixed at $\mu=-10t$. The total number of particles $N$ is about $18,218$. For the second scenario (lower panels in Fig. ~\ref{Mott1}), the annular trap depth $V_a=20t$,  $U=25t$,  the attractive trench potential (cf., Eqs. (\ref{rect-gauss-pot}) \textit{and} (\ref{trench-addon-pot}))   $V_b=-13t$, $d_0=10$, chemical potential $\mu=-8.2t$, and the total number of particles $N$ is about $7082$. The panels at the far left (Figs. \ref{Mott1}$(a)$ and $(e)$) show false color plots of  the absolute value of superfluid density ($|\langle{b_j\rangle}|$)  , while Figs. \ref{Mott1}$(b)$ and $(f)$ show false color plots of the number density profile $\langle{n_j\rangle}$, all for the case with no superflow. The four panels at right show the current distributions for  cases with superflow, with $l=6$ for the first scenario (upper panels) and $l=4$ for the second scenario (lower panels). In both the cases, the temperature is taken to be $T=t$. We find that  the entropy per particle is  lower ($S/N=0.138k_B$) for the first scenario than for the second scenario ($S/N=0.532k_B$).

\begin{figure}[t!]
\includegraphics[width=9cm]{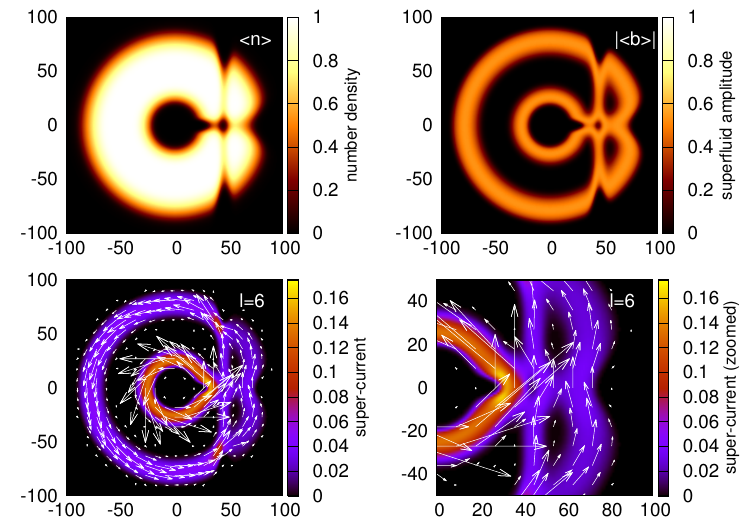}
	\caption{(Color online) Configurations in the case with two different repulsive barriers, one radially across the annulus ($V_b^{(1)}=6t$; cf. Eq. (\ref{rect-gauss-pot})), and the other perpendicular to the radial barrier ($V_b^{(2)}=6t$; cf. Eq. (\ref{perp-barr-pot})). The various  quantities shown are the same ones as in Fig. ~\ref{Mott1}. }
\label{Mott2}
\end{figure}


Another interesting configuration with a singly occupied Mott phase can be generated  by adding  a second repulsive barrier (cf., Eq. (\ref{perp-barr-pot})) of the same height ($V_b^{(2)}=6t$) perpendicular to the barrier corresponding to Eq. (\ref{rect-gauss-pot}) which is radially across the annulus ($V_b^{(1)}=6t$). In this case, a four way junction can be formed as shown in Fig. \ref{Mott2}. The other parameters are the same as in the upper panel of Fig. \ref{Mott1}. The total number of particles $N$ is about $17,256$ and the entropy per particle is $S/N=0.147k_B$.

\begin{figure}[b]
\includegraphics[width=9cm]{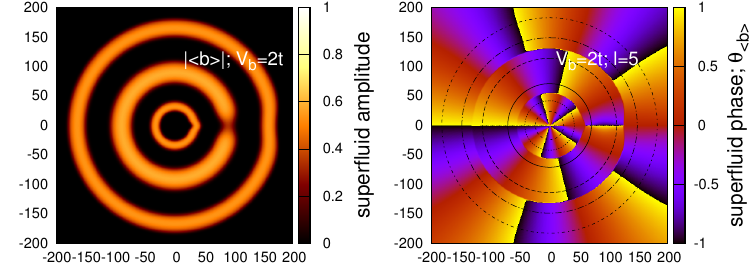}
\caption{(Color online) \textit{Two} $\langle{n\rangle}=1$ concentric Mott regions separating \textit{three} concentric superfluid regions. The left panel shows a false color plot of the superfluid order parameter. The right panel shows a false color plot of the phase of the superfluid order parameter which is defined in the range from $-\pi$ to $\pi$. The figures clearly show the existence of phase slips between the central and the other regions; we find that the  threshold  winding number  for the occurrence of phase slips is $l=4$ for $V_b=2t$.}
\label{Mott3}
\end{figure}

The superfluid channels in the various configurations above give rise to different critical currents in different parts of the circuit. Phase slips can hence occur differently in different parts of the circuit at sufficiently high currents and give rise to interesting current configurations. We believe such effects portend new possibilities of value in the design of atomtronic circuits.

\subsection{Configurations with two $\langle{n\rangle}=1$ Mott regions}
Even more complex configurations can be generated with Mott regions numbering more than one. The next level of complexity is to allow two concentric $\langle{n\rangle}=1$ Mott regions separated by a superfluid region (of $\langle{n\rangle}> 1$), with two other concentric superfluid regions on either side, as shown in Fig. \ref{Mott3}.  Varying the strength of the repulsive barrier leads to different currents, corresponding to different phase winding numbers, in each of three superfluid rings. As shown in Fig. \ref{Mott3}, for $V_b=2t$ all the three rings remain separate but a superfluid constriction is formed in the middle ring. Phase slip occurs for winding numbers more than the critical winding number $l_{max}=4$ for the middle ring. The annular trap depth is $V_a=90t$ and an angular-Gaussian barrier ($V_B$) is used of width $\theta_0=0.2$, the rest of the parameters being the same as in Fig. \ref{Mott1} left panel. The total number of particles in this case is $N=87,533$ and the entropy per particle is $S/N=0.073k_B$. For configurations with several superfluid rings, the repulsive barrier can give rise to either superfluid constrictions or indentations of the superfluid regions depending on their positions in the annular trap. For an odd number of superfluid rings, the middle ring can be used as a SQUID device. The indentations give rise to reduction of superfluid current in a given ring.

In the above two sections we have demonstrated different circuits of bosonic superfluid based on a lattice. Addition of Mott phases by increasing $V_a$ as well as creating {\it barrier} or {\it trench} potentials of various strengths ($V_b$) leads to formation and manipulation of ``wires" for atomtronic circuits. These `persistent current' based atomtronic circuits can be used for AQUID (Atomtronic Quantum Interference Device) applications \cite{PhysRevA.91.033607, PhysRevA.99.043613, 10.21468/SciPostPhys.11.4.080} or quantum qubit implimentations \cite{Aghamalyan_2015, Aghamalyan_2016}. We believe these circuits can be implimented in the experiments based on the current technological advancements. 

\section{Concluding comments}
In this paper, we have explored persistent current and phase slip phenomena in cold Bosonic atom superfluids in annular traps with barriers that are additionally subjected to an optical lattice potential. For this purpose, we modeled the system in terms of a BHM on a square lattice in the presence of external (annular trap and barrier) potentials, and extended the strong-coupling expansion to steady states that have superflow. We have reported quantitative studies of the phase transitions between persistent current states with different quantised circulations, resulting from varying the temperature and the barrier height. Furthermore, we have explored the new and complex superflow configurations that result if the parameters of the system are adjusted  to permit Mott regions inside the trap. These provide possibilities for the creation of new atomtronic circuits. In particular, the presence of Mott phases in these devices creates an additional way to tune the superfluid and insulating regions of the device,  perhaps with more tunability than in solid-state devices. We hope these ideas will be explored in experimental realizations soon.
\vspace{0.4cm}

\section{Acknowledgements}
MG would like to acknowledge financial support from IUSSTF (Indo-US Science and Technology forum). HRK gratefully acknowledges support from the Science and Engineering Research Board of the Department of Science and Technology, India, under grants No. SR/S2/JCB-19/2005 and SB/DF/005/2017, and the Indian National Science Academy under grant number No. INSA/SP/SS/2023/. Work at Georgetown University was supported by the U.S. Department of Energy (DOE), Office of Science, Basic Energy Sciences (BES) under Award DE-FG02-08ER46542.

 
 \appendix*
  \section{Current conservation for mean-field $J_{ij}$}
  \label{appA}
    
The mean-field current density from site $j$ to $j'$ obtained from Eq. (\ref{eqJ})  is given by
\begin{equation}
\langle{J_{jj'}\rangle}=-i [ \langle{b_j\rangle}^*_0 t_{jj'} \langle{b_{j'}\rangle}_0 - \langle{b_{j'}\rangle}^*_0 t_{j'j} \langle{b_j\rangle}_0 ]
\end{equation}


Current conservation would require that the sum of outgoing (mean-field) currents must equal to the sum of ingoing (mean-field) currents at every site. In other words we require
\begin{equation}
\sum_{j'}\langle{J_{jj'}\rangle} = 0
\end{equation}

Using the mean-field self-consistency relation (see Section IIIA)
$\phi_j \equiv \sum_{j'}t_{jj'} \langle b_{j'} \rangle_0$   we can write
\begin{eqnarray}
\sum_{j'} \langle{J_{jj'}\rangle}&&=-i [\langle{b_j\rangle}^*_0 \phi_j - \phi^*_j \langle{b_j\rangle}_0 ] \\
&&=i[\phi_j \frac{\partial}{\partial \phi_j}\tilde{f}_j-\phi^*_j\frac{\partial}{\partial \phi_j^*}\tilde{f}_j]
\end{eqnarray}
where  $\tilde{f}_j \equiv -k_B T \ln \tilde{z}_j$ is the free energy at temperature $T$ for 
$\tilde{\mathcal{H}}_{0j}$, the mean field Hamiltonian at site $j$, with $\tilde{z}_j$ being the corresponding partition function (cf., Eq.s \ref{eqH0} and \ref{eqbjzj}).  

It is easy to see that $\tilde{f}_j$ can only be a function of $|\phi_j|^2$ or $\phi_j \phi_j^*$, i.e., it is gauge invariant with respect to an arbitrary change of phase of $\phi_j$, because the phase of 
$\phi_j$ can be absorbed into a (unitary) redefinition of $b^{\dagger}_j$. Hence, 
$\frac{\partial}{\partial \phi_j}\tilde{f}_j=\phi_j^*\frac{\partial}{\partial |\phi_j|^2}\tilde{f}_j$ and $\frac{\partial}{\partial \phi_j^*}\tilde{f}_j=\phi\frac{\partial}{\partial |\phi_j|^2}\tilde{f}_j$, from which it follows that 
\begin{equation}
\sum_{j'}\langle{J_{jj'}\rangle} = i[\phi_j\phi_j^*-\phi_j^*\phi_j]\frac{\partial}{\partial |\phi_j|^2}\tilde{f}_j = 0.
\end{equation}

\bibliography{annuler.bib}

\begin{thebibliography}{43}%
\makeatletter
\providecommand \@ifxundefined [1]{%
 \@ifx{#1\undefined}
}%
\providecommand \@ifnum [1]{%
 \ifnum #1\expandafter \@firstoftwo
 \else \expandafter \@secondoftwo
 \fi
}%
\providecommand \@ifx [1]{%
 \ifx #1\expandafter \@firstoftwo
 \else \expandafter \@secondoftwo
 \fi
}%
\providecommand \natexlab [1]{#1}%
\providecommand \enquote  [1]{``#1''}%
\providecommand \bibnamefont  [1]{#1}%
\providecommand \bibfnamefont [1]{#1}%
\providecommand \citenamefont [1]{#1}%
\providecommand \href@noop [0]{\@secondoftwo}%
\providecommand \href [0]{\begingroup \@sanitize@url \@href}%
\providecommand \@href[1]{\@@startlink{#1}\@@href}%
\providecommand \@@href[1]{\endgroup#1\@@endlink}%
\providecommand \@sanitize@url [0]{\catcode `\\12\catcode `\$12\catcode
  `\&12\catcode `\#12\catcode `\^12\catcode `\_12\catcode `\%12\relax}%
\providecommand \@@startlink[1]{}%
\providecommand \@@endlink[0]{}%
\providecommand \url  [0]{\begingroup\@sanitize@url \@url }%
\providecommand \@url [1]{\endgroup\@href {#1}{\urlprefix }}%
\providecommand \urlprefix  [0]{URL }%
\providecommand \Eprint [0]{\href }%
\providecommand \doibase [0]{https://doi.org/}%
\providecommand \selectlanguage [0]{\@gobble}%
\providecommand \bibinfo  [0]{\@secondoftwo}%
\providecommand \bibfield  [0]{\@secondoftwo}%
\providecommand \translation [1]{[#1]}%
\providecommand \BibitemOpen [0]{}%
\providecommand \bibitemStop [0]{}%
\providecommand \bibitemNoStop [0]{.\EOS\space}%
\providecommand \EOS [0]{\spacefactor3000\relax}%
\providecommand \BibitemShut  [1]{\csname bibitem#1\endcsname}%
\let\auto@bib@innerbib\@empty
\bibitem [{\citenamefont {Seaman}\ \emph {et~al.}(2007)\citenamefont {Seaman},
  \citenamefont {Kr\"amer}, \citenamefont {Anderson},\ and\ \citenamefont
  {Holland}}]{Atomtronics}%
  \BibitemOpen
  \bibfield  {author} {\bibinfo {author} {\bibfnamefont {B.~T.}\ \bibnamefont
  {Seaman}}, \bibinfo {author} {\bibfnamefont {M.}~\bibnamefont {Kr\"amer}},
  \bibinfo {author} {\bibfnamefont {D.~Z.}\ \bibnamefont {Anderson}},\ and\
  \bibinfo {author} {\bibfnamefont {M.~J.}\ \bibnamefont {Holland}},\
  }\bibfield  {title} {\bibinfo {title} {Atomtronics: Ultracold-atom analogs of
  electronic devices},\ }\href {https://doi.org/10.1103/PhysRevA.75.023615}
  {\bibfield  {journal} {\bibinfo  {journal} {Phys. Rev. A}\ }\textbf {\bibinfo
  {volume} {75}},\ \bibinfo {pages} {023615} (\bibinfo {year}
  {2007})}\BibitemShut {NoStop}%
\bibitem [{\citenamefont {Amico}\ \emph {et~al.}(2022)\citenamefont {Amico},
  \citenamefont {Anderson}, \citenamefont {Boshier}, \citenamefont {Brantut},
  \citenamefont {Kwek}, \citenamefont {Minguzzi},\ and\ \citenamefont {von
  Klitzing}}]{RevModPhys.94.041001}%
  \BibitemOpen
  \bibfield  {author} {\bibinfo {author} {\bibfnamefont {L.}~\bibnamefont
  {Amico}}, \bibinfo {author} {\bibfnamefont {D.}~\bibnamefont {Anderson}},
  \bibinfo {author} {\bibfnamefont {M.}~\bibnamefont {Boshier}}, \bibinfo
  {author} {\bibfnamefont {J.-P.}\ \bibnamefont {Brantut}}, \bibinfo {author}
  {\bibfnamefont {L.-C.}\ \bibnamefont {Kwek}}, \bibinfo {author}
  {\bibfnamefont {A.}~\bibnamefont {Minguzzi}},\ and\ \bibinfo {author}
  {\bibfnamefont {W.}~\bibnamefont {von Klitzing}},\ }\bibfield  {title}
  {\bibinfo {title} {Colloquium: Atomtronic circuits: From many-body physics to
  quantum technologies},\ }\href {https://doi.org/10.1103/RevModPhys.94.041001}
  {\bibfield  {journal} {\bibinfo  {journal} {Rev. Mod. Phys.}\ }\textbf
  {\bibinfo {volume} {94}},\ \bibinfo {pages} {041001} (\bibinfo {year}
  {2022})}\BibitemShut {NoStop}%
\bibitem [{\citenamefont {Pepino}\ \emph {et~al.}(2009)\citenamefont {Pepino},
  \citenamefont {Cooper}, \citenamefont {Anderson},\ and\ \citenamefont
  {Holland}}]{DandT}%
  \BibitemOpen
  \bibfield  {author} {\bibinfo {author} {\bibfnamefont {R.~A.}\ \bibnamefont
  {Pepino}}, \bibinfo {author} {\bibfnamefont {J.}~\bibnamefont {Cooper}},
  \bibinfo {author} {\bibfnamefont {D.~Z.}\ \bibnamefont {Anderson}},\ and\
  \bibinfo {author} {\bibfnamefont {M.~J.}\ \bibnamefont {Holland}},\
  }\bibfield  {title} {\bibinfo {title} {Atomtronic circuits of diodes and
  transistors},\ }\href {https://doi.org/10.1103/PhysRevLett.103.140405}
  {\bibfield  {journal} {\bibinfo  {journal} {Phys. Rev. Lett.}\ }\textbf
  {\bibinfo {volume} {103}},\ \bibinfo {pages} {140405} (\bibinfo {year}
  {2009})}\BibitemShut {NoStop}%
\bibitem [{\citenamefont {Jendrzejewski}\ \emph {et~al.}(2014)\citenamefont
  {Jendrzejewski}, \citenamefont {Eckel}, \citenamefont {Murray}, \citenamefont
  {Lanier}, \citenamefont {Edwards}, \citenamefont {Lobb},\ and\ \citenamefont
  {Campbell}}]{resistance}%
  \BibitemOpen
  \bibfield  {author} {\bibinfo {author} {\bibfnamefont {F.}~\bibnamefont
  {Jendrzejewski}}, \bibinfo {author} {\bibfnamefont {S.}~\bibnamefont
  {Eckel}}, \bibinfo {author} {\bibfnamefont {N.}~\bibnamefont {Murray}},
  \bibinfo {author} {\bibfnamefont {C.}~\bibnamefont {Lanier}}, \bibinfo
  {author} {\bibfnamefont {M.}~\bibnamefont {Edwards}}, \bibinfo {author}
  {\bibfnamefont {C.~J.}\ \bibnamefont {Lobb}},\ and\ \bibinfo {author}
  {\bibfnamefont {G.~K.}\ \bibnamefont {Campbell}},\ }\bibfield  {title}
  {\bibinfo {title} {Resistive flow in a weakly interacting bose-einstein
  condensate},\ }\href {https://doi.org/10.1103/PhysRevLett.113.045305}
  {\bibfield  {journal} {\bibinfo  {journal} {Phys. Rev. Lett.}\ }\textbf
  {\bibinfo {volume} {113}},\ \bibinfo {pages} {045305} (\bibinfo {year}
  {2014})}\BibitemShut {NoStop}%
\bibitem [{\citenamefont {Eckel}\ \emph
  {et~al.}(2014{\natexlab{a}})\citenamefont {Eckel}, \citenamefont {Lee},
  \citenamefont {Jendrzejewski}, \citenamefont {Murray}, \citenamefont {Clark},
  \citenamefont {Lobb}, \citenamefont {Phillips}, \citenamefont {Edwards},\
  and\ \citenamefont {Campbell}}]{Hysteresis}%
  \BibitemOpen
  \bibfield  {author} {\bibinfo {author} {\bibfnamefont {S.}~\bibnamefont
  {Eckel}}, \bibinfo {author} {\bibfnamefont {J.~G.}\ \bibnamefont {Lee}},
  \bibinfo {author} {\bibfnamefont {F.}~\bibnamefont {Jendrzejewski}}, \bibinfo
  {author} {\bibfnamefont {N.}~\bibnamefont {Murray}}, \bibinfo {author}
  {\bibfnamefont {C.~W.}\ \bibnamefont {Clark}}, \bibinfo {author}
  {\bibfnamefont {C.~J.}\ \bibnamefont {Lobb}}, \bibinfo {author}
  {\bibfnamefont {W.~D.}\ \bibnamefont {Phillips}}, \bibinfo {author}
  {\bibfnamefont {M.}~\bibnamefont {Edwards}},\ and\ \bibinfo {author}
  {\bibfnamefont {G.~K.}\ \bibnamefont {Campbell}},\ }\bibfield  {title}
  {\bibinfo {title} {Hysteresis in a quantized superfluid `atomtronic'
  circuit},\ }\href {https://doi.org/10.1038/nature12958} {\bibfield  {journal}
  {\bibinfo  {journal} {Nature}\ }\textbf {\bibinfo {volume} {506}},\ \bibinfo
  {pages} {200} (\bibinfo {year} {2014}{\natexlab{a}})}\BibitemShut {NoStop}%
\bibitem [{\citenamefont {Amico}\ \emph {et~al.}(2014)\citenamefont {Amico},
  \citenamefont {Aghamalyan}, \citenamefont {Auksztol}, \citenamefont {Crepaz},
  \citenamefont {Dumke},\ and\ \citenamefont {Kwek}}]{Squbit}%
  \BibitemOpen
  \bibfield  {author} {\bibinfo {author} {\bibfnamefont {L.}~\bibnamefont
  {Amico}}, \bibinfo {author} {\bibfnamefont {D.}~\bibnamefont {Aghamalyan}},
  \bibinfo {author} {\bibfnamefont {F.}~\bibnamefont {Auksztol}}, \bibinfo
  {author} {\bibfnamefont {H.}~\bibnamefont {Crepaz}}, \bibinfo {author}
  {\bibfnamefont {R.}~\bibnamefont {Dumke}},\ and\ \bibinfo {author}
  {\bibfnamefont {L.~C.}\ \bibnamefont {Kwek}},\ }\bibfield  {title} {\bibinfo
  {title} {Superfluid qubit systems with ring shaped optical lattices},\
  }\bibfield  {journal} {\bibinfo  {journal} {Sci. Rep.}\ }\textbf {\bibinfo
  {volume} {4}},\ \href {https://doi.org/10.1038/srep04298} {10.1038/srep04298}
  (\bibinfo {year} {2014})\BibitemShut {NoStop}%
\bibitem [{\citenamefont {Wright}\ \emph
  {et~al.}(2013{\natexlab{a}})\citenamefont {Wright}, \citenamefont
  {Blakestad}, \citenamefont {Lobb}, \citenamefont {Phillips},\ and\
  \citenamefont {Campbell}}]{GCampbell2013}%
  \BibitemOpen
  \bibfield  {author} {\bibinfo {author} {\bibfnamefont {K.~C.}\ \bibnamefont
  {Wright}}, \bibinfo {author} {\bibfnamefont {R.~B.}\ \bibnamefont
  {Blakestad}}, \bibinfo {author} {\bibfnamefont {C.~J.}\ \bibnamefont {Lobb}},
  \bibinfo {author} {\bibfnamefont {W.~D.}\ \bibnamefont {Phillips}},\ and\
  \bibinfo {author} {\bibfnamefont {G.~K.}\ \bibnamefont {Campbell}},\
  }\bibfield  {title} {\bibinfo {title} {Driving phase slips in a superfluid
  atom circuit with a rotating weak link},\ }\href
  {https://doi.org/10.1103/PhysRevLett.110.025302} {\bibfield  {journal}
  {\bibinfo  {journal} {Phys. Rev. Lett.}\ }\textbf {\bibinfo {volume} {110}},\
  \bibinfo {pages} {025302} (\bibinfo {year} {2013}{\natexlab{a}})}\BibitemShut
  {NoStop}%
\bibitem [{\citenamefont {Ramanathan}\ \emph {et~al.}(2011)\citenamefont
  {Ramanathan}, \citenamefont {Wright}, \citenamefont {Muniz}, \citenamefont
  {Zelan}, \citenamefont {Hill}, \citenamefont {Lobb}, \citenamefont
  {Helmerson}, \citenamefont {Phillips},\ and\ \citenamefont
  {Campbell}}]{GCampbell2011}%
  \BibitemOpen
  \bibfield  {author} {\bibinfo {author} {\bibfnamefont {A.}~\bibnamefont
  {Ramanathan}}, \bibinfo {author} {\bibfnamefont {K.~C.}\ \bibnamefont
  {Wright}}, \bibinfo {author} {\bibfnamefont {S.~R.}\ \bibnamefont {Muniz}},
  \bibinfo {author} {\bibfnamefont {M.}~\bibnamefont {Zelan}}, \bibinfo
  {author} {\bibfnamefont {W.~T.}\ \bibnamefont {Hill}}, \bibinfo {author}
  {\bibfnamefont {C.~J.}\ \bibnamefont {Lobb}}, \bibinfo {author}
  {\bibfnamefont {K.}~\bibnamefont {Helmerson}}, \bibinfo {author}
  {\bibfnamefont {W.~D.}\ \bibnamefont {Phillips}},\ and\ \bibinfo {author}
  {\bibfnamefont {G.~K.}\ \bibnamefont {Campbell}},\ }\bibfield  {title}
  {\bibinfo {title} {Superflow in a toroidal bose-einstein condensate: An atom
  circuit with a tunable weak link},\ }\href
  {https://doi.org/10.1103/PhysRevLett.106.130401} {\bibfield  {journal}
  {\bibinfo  {journal} {Phys. Rev. Lett.}\ }\textbf {\bibinfo {volume} {106}},\
  \bibinfo {pages} {130401} (\bibinfo {year} {2011})}\BibitemShut {NoStop}%
\bibitem [{\citenamefont {Ryu}\ \emph {et~al.}(2007)\citenamefont {Ryu},
  \citenamefont {Andersen}, \citenamefont {Clad\'e}, \citenamefont {Natarajan},
  \citenamefont {Helmerson},\ and\ \citenamefont {Phillips}}]{CRyu}%
  \BibitemOpen
  \bibfield  {author} {\bibinfo {author} {\bibfnamefont {C.}~\bibnamefont
  {Ryu}}, \bibinfo {author} {\bibfnamefont {M.~F.}\ \bibnamefont {Andersen}},
  \bibinfo {author} {\bibfnamefont {P.}~\bibnamefont {Clad\'e}}, \bibinfo
  {author} {\bibfnamefont {V.}~\bibnamefont {Natarajan}}, \bibinfo {author}
  {\bibfnamefont {K.}~\bibnamefont {Helmerson}},\ and\ \bibinfo {author}
  {\bibfnamefont {W.~D.}\ \bibnamefont {Phillips}},\ }\bibfield  {title}
  {\bibinfo {title} {Observation of persistent flow of a bose-einstein
  condensate in a toroidal trap},\ }\href
  {https://doi.org/10.1103/PhysRevLett.99.260401} {\bibfield  {journal}
  {\bibinfo  {journal} {Phys. Rev. Lett.}\ }\textbf {\bibinfo {volume} {99}},\
  \bibinfo {pages} {260401} (\bibinfo {year} {2007})}\BibitemShut {NoStop}%
\bibitem [{\citenamefont {Wright}\ \emph
  {et~al.}(2013{\natexlab{b}})\citenamefont {Wright}, \citenamefont
  {Blakestad}, \citenamefont {Lobb}, \citenamefont {Phillips},\ and\
  \citenamefont {Campbell}}]{StirredSup}%
  \BibitemOpen
  \bibfield  {author} {\bibinfo {author} {\bibfnamefont {K.~C.}\ \bibnamefont
  {Wright}}, \bibinfo {author} {\bibfnamefont {R.~B.}\ \bibnamefont
  {Blakestad}}, \bibinfo {author} {\bibfnamefont {C.~J.}\ \bibnamefont {Lobb}},
  \bibinfo {author} {\bibfnamefont {W.~D.}\ \bibnamefont {Phillips}},\ and\
  \bibinfo {author} {\bibfnamefont {G.~K.}\ \bibnamefont {Campbell}},\
  }\bibfield  {title} {\bibinfo {title} {Threshold for creating excitations in
  a stirred superfluid ring},\ }\href
  {https://doi.org/10.1103/PhysRevA.88.063633} {\bibfield  {journal} {\bibinfo
  {journal} {Phys. Rev. A}\ }\textbf {\bibinfo {volume} {88}},\ \bibinfo
  {pages} {063633} (\bibinfo {year} {2013}{\natexlab{b}})}\BibitemShut
  {NoStop}%
\bibitem [{\citenamefont {Murray}\ \emph {et~al.}(2013)\citenamefont {Murray},
  \citenamefont {Krygier}, \citenamefont {Edwards}, \citenamefont {Wright},
  \citenamefont {Campbell},\ and\ \citenamefont {Clark}}]{Murray}%
  \BibitemOpen
  \bibfield  {author} {\bibinfo {author} {\bibfnamefont {N.}~\bibnamefont
  {Murray}}, \bibinfo {author} {\bibfnamefont {M.}~\bibnamefont {Krygier}},
  \bibinfo {author} {\bibfnamefont {M.}~\bibnamefont {Edwards}}, \bibinfo
  {author} {\bibfnamefont {K.~C.}\ \bibnamefont {Wright}}, \bibinfo {author}
  {\bibfnamefont {G.~K.}\ \bibnamefont {Campbell}},\ and\ \bibinfo {author}
  {\bibfnamefont {C.~W.}\ \bibnamefont {Clark}},\ }\bibfield  {title} {\bibinfo
  {title} {Probing the circulation of ring-shaped bose-einstein condensates},\
  }\href {https://doi.org/10.1103/PhysRevA.88.053615} {\bibfield  {journal}
  {\bibinfo  {journal} {Phys. Rev. A}\ }\textbf {\bibinfo {volume} {88}},\
  \bibinfo {pages} {053615} (\bibinfo {year} {2013})}\BibitemShut {NoStop}%
\bibitem [{\citenamefont {Polo}\ \emph {et~al.}(2023)\citenamefont {Polo},
  \citenamefont {Chetcuti}, \citenamefont {Domanti}, \citenamefont {Kitson},
  \citenamefont {Osterloh}, \citenamefont {Perciavalle}, \citenamefont
  {Singh},\ and\ \citenamefont {Amico}}]{polo2023perspective}%
  \BibitemOpen
  \bibfield  {author} {\bibinfo {author} {\bibfnamefont {J.}~\bibnamefont
  {Polo}}, \bibinfo {author} {\bibfnamefont {W.~J.}\ \bibnamefont {Chetcuti}},
  \bibinfo {author} {\bibfnamefont {E.~C.}\ \bibnamefont {Domanti}}, \bibinfo
  {author} {\bibfnamefont {P.}~\bibnamefont {Kitson}}, \bibinfo {author}
  {\bibfnamefont {A.}~\bibnamefont {Osterloh}}, \bibinfo {author}
  {\bibfnamefont {F.}~\bibnamefont {Perciavalle}}, \bibinfo {author}
  {\bibfnamefont {V.~P.}\ \bibnamefont {Singh}},\ and\ \bibinfo {author}
  {\bibfnamefont {L.}~\bibnamefont {Amico}},\ }\href@noop {} {\bibinfo {title}
  {Perspective on new implementations of atomtronic circuits}} (\bibinfo {year}
  {2023}),\ \Eprint {https://arxiv.org/abs/2311.16257} {arXiv:2311.16257
  [cond-mat.quant-gas]} \BibitemShut {NoStop}%
\bibitem [{\citenamefont {Pezzè}\ \emph {et~al.}(2024)\citenamefont {Pezzè},
  \citenamefont {Xhani}, \citenamefont {Daix}, \citenamefont {Grani},
  \citenamefont {Donelli}, \citenamefont {Scazza}, \citenamefont
  {Hernandez-Rajkov}, \citenamefont {Kwon}, \citenamefont {Del~Pace},\ and\
  \citenamefont {Roati}}]{Pezzè2024}%
  \BibitemOpen
  \bibfield  {author} {\bibinfo {author} {\bibfnamefont {L.}~\bibnamefont
  {Pezzè}}, \bibinfo {author} {\bibfnamefont {K.}~\bibnamefont {Xhani}},
  \bibinfo {author} {\bibfnamefont {C.}~\bibnamefont {Daix}}, \bibinfo {author}
  {\bibfnamefont {N.}~\bibnamefont {Grani}}, \bibinfo {author} {\bibfnamefont
  {B.}~\bibnamefont {Donelli}}, \bibinfo {author} {\bibfnamefont
  {F.}~\bibnamefont {Scazza}}, \bibinfo {author} {\bibfnamefont
  {D.}~\bibnamefont {Hernandez-Rajkov}}, \bibinfo {author} {\bibfnamefont
  {W.~J.}\ \bibnamefont {Kwon}}, \bibinfo {author} {\bibfnamefont
  {G.}~\bibnamefont {Del~Pace}},\ and\ \bibinfo {author} {\bibfnamefont
  {G.}~\bibnamefont {Roati}},\ }\bibfield  {title} {\bibinfo {title}
  {Stabilizing persistent currents in an atomtronic josephson junction
  necklace},\ }\href@noop {} {\bibfield  {journal} {\bibinfo  {journal} {Nature
  Communications}\ }\textbf {\bibinfo {volume} {15}},\ \bibinfo {pages} {4831}
  (\bibinfo {year} {2024})}\BibitemShut {NoStop}%
\bibitem [{\citenamefont {Eckel}\ \emph
  {et~al.}(2014{\natexlab{b}})\citenamefont {Eckel}, \citenamefont
  {Jendrzejewski}, \citenamefont {Kumar}, \citenamefont {Lobb},\ and\
  \citenamefont {Campbell}}]{CurrPhase}%
  \BibitemOpen
  \bibfield  {author} {\bibinfo {author} {\bibfnamefont {S.}~\bibnamefont
  {Eckel}}, \bibinfo {author} {\bibfnamefont {F.}~\bibnamefont
  {Jendrzejewski}}, \bibinfo {author} {\bibfnamefont {A.}~\bibnamefont
  {Kumar}}, \bibinfo {author} {\bibfnamefont {C.~J.}\ \bibnamefont {Lobb}},\
  and\ \bibinfo {author} {\bibfnamefont {G.~K.}\ \bibnamefont {Campbell}},\
  }\bibfield  {title} {\bibinfo {title} {Interferometric measurement of the
  current-phase relationship of a superfluid weak link},\ }\href
  {https://doi.org/10.1103/PhysRevX.4.031052} {\bibfield  {journal} {\bibinfo
  {journal} {Phys. Rev. X}\ }\textbf {\bibinfo {volume} {4}},\ \bibinfo {pages}
  {031052} (\bibinfo {year} {2014}{\natexlab{b}})}\BibitemShut {NoStop}%
\bibitem [{\citenamefont {Piazza}\ \emph {et~al.}(2010)\citenamefont {Piazza},
  \citenamefont {Collins},\ and\ \citenamefont {Smerzi}}]{Piazza}%
  \BibitemOpen
  \bibfield  {author} {\bibinfo {author} {\bibfnamefont {F.}~\bibnamefont
  {Piazza}}, \bibinfo {author} {\bibfnamefont {L.~A.}\ \bibnamefont
  {Collins}},\ and\ \bibinfo {author} {\bibfnamefont {A.}~\bibnamefont
  {Smerzi}},\ }\bibfield  {title} {\bibinfo {title} {Current-phase relation of
  a bose-einstein condensate flowing through a weak link},\ }\href
  {https://doi.org/10.1103/PhysRevA.81.033613} {\bibfield  {journal} {\bibinfo
  {journal} {Phys. Rev. A}\ }\textbf {\bibinfo {volume} {81}},\ \bibinfo
  {pages} {033613} (\bibinfo {year} {2010})}\BibitemShut {NoStop}%
\bibitem [{\citenamefont {Piazza}\ \emph {et~al.}(2009)\citenamefont {Piazza},
  \citenamefont {Collins},\ and\ \citenamefont {Smerzi}}]{Piazza2}%
  \BibitemOpen
  \bibfield  {author} {\bibinfo {author} {\bibfnamefont {F.}~\bibnamefont
  {Piazza}}, \bibinfo {author} {\bibfnamefont {L.~A.}\ \bibnamefont
  {Collins}},\ and\ \bibinfo {author} {\bibfnamefont {A.}~\bibnamefont
  {Smerzi}},\ }\bibfield  {title} {\bibinfo {title} {Vortex-induced phase-slip
  dissipation in a toroidal bose-einstein condensate flowing through a
  barrier},\ }\href {https://doi.org/10.1103/PhysRevA.80.021601} {\bibfield
  {journal} {\bibinfo  {journal} {Phys. Rev. A}\ }\textbf {\bibinfo {volume}
  {80}},\ \bibinfo {pages} {021601} (\bibinfo {year} {2009})}\BibitemShut
  {NoStop}%
\bibitem [{\citenamefont {Piazza}\ \emph {et~al.}(2013)\citenamefont {Piazza},
  \citenamefont {Collins},\ and\ \citenamefont {Smerzi}}]{Piazza3}%
  \BibitemOpen
  \bibfield  {author} {\bibinfo {author} {\bibfnamefont {F.}~\bibnamefont
  {Piazza}}, \bibinfo {author} {\bibfnamefont {L.~A.}\ \bibnamefont
  {Collins}},\ and\ \bibinfo {author} {\bibfnamefont {A.}~\bibnamefont
  {Smerzi}},\ }\bibfield  {title} {\bibinfo {title} {Critical velocity for a
  toroidal bose–einstein condensate flowing through a barrier},\ }\href
  {https://doi.org/10.1088/0953-4075/46/9/095302} {\bibfield  {journal}
  {\bibinfo  {journal} {Journal of Physics B: Atomic, Molecular and Optical
  Physics}\ }\textbf {\bibinfo {volume} {46}},\ \bibinfo {pages} {095302}
  (\bibinfo {year} {2013})}\BibitemShut {NoStop}%
\bibitem [{\citenamefont {Gallemí}\ \emph {et~al.}(2015)\citenamefont
  {Gallemí}, \citenamefont {Mateo}, \citenamefont {Mayol},\ and\ \citenamefont
  {Guilleumas}}]{Gallemi2016}%
  \BibitemOpen
  \bibfield  {author} {\bibinfo {author} {\bibfnamefont {A.}~\bibnamefont
  {Gallemí}}, \bibinfo {author} {\bibfnamefont {A.~M.}\ \bibnamefont {Mateo}},
  \bibinfo {author} {\bibfnamefont {R.}~\bibnamefont {Mayol}},\ and\ \bibinfo
  {author} {\bibfnamefont {M.}~\bibnamefont {Guilleumas}},\ }\bibfield  {title}
  {\bibinfo {title} {Coherent quantum phase slip in two-component bosonic
  atomtronic circuits},\ }\href {https://doi.org/10.1088/1367-2630/18/1/015003}
  {\bibfield  {journal} {\bibinfo  {journal} {New Journal of Physics}\ }\textbf
  {\bibinfo {volume} {18}},\ \bibinfo {pages} {015003} (\bibinfo {year}
  {2015})}\BibitemShut {NoStop}%
\bibitem [{\citenamefont {Polo}\ \emph {et~al.}(2019)\citenamefont {Polo},
  \citenamefont {Dubessy}, \citenamefont {Pedri}, \citenamefont {Perrin},\ and\
  \citenamefont {Minguzzi}}]{1DringGasDecay}%
  \BibitemOpen
  \bibfield  {author} {\bibinfo {author} {\bibfnamefont {J.}~\bibnamefont
  {Polo}}, \bibinfo {author} {\bibfnamefont {R.}~\bibnamefont {Dubessy}},
  \bibinfo {author} {\bibfnamefont {P.}~\bibnamefont {Pedri}}, \bibinfo
  {author} {\bibfnamefont {H.}~\bibnamefont {Perrin}},\ and\ \bibinfo {author}
  {\bibfnamefont {A.}~\bibnamefont {Minguzzi}},\ }\bibfield  {title} {\bibinfo
  {title} {Oscillations and decay of superfluid currents in a one-dimensional
  bose gas on a ring},\ }\href {https://doi.org/10.1103/PhysRevLett.123.195301}
  {\bibfield  {journal} {\bibinfo  {journal} {Phys. Rev. Lett.}\ }\textbf
  {\bibinfo {volume} {123}},\ \bibinfo {pages} {195301} (\bibinfo {year}
  {2019})}\BibitemShut {NoStop}%
\bibitem [{\citenamefont {Andriati}\ and\ \citenamefont
  {Gammal}(2019)}]{FewBAnn}%
  \BibitemOpen
  \bibfield  {author} {\bibinfo {author} {\bibfnamefont {A.~V.}\ \bibnamefont
  {Andriati}}\ and\ \bibinfo {author} {\bibfnamefont {A.}~\bibnamefont
  {Gammal}},\ }\bibfield  {title} {\bibinfo {title} {Superfluid fraction of few
  bosons in an annular geometry in the presence of a rotating weak link},\
  }\href {https://doi.org/10.1103/PhysRevA.100.063625} {\bibfield  {journal}
  {\bibinfo  {journal} {Phys. Rev. A}\ }\textbf {\bibinfo {volume} {100}},\
  \bibinfo {pages} {063625} (\bibinfo {year} {2019})}\BibitemShut {NoStop}%
\bibitem [{\citenamefont {Kiehn}\ \emph {et~al.}(2022)\citenamefont {Kiehn},
  \citenamefont {Singh},\ and\ \citenamefont {Mathey}}]{asquidTorr}%
  \BibitemOpen
  \bibfield  {author} {\bibinfo {author} {\bibfnamefont {H.}~\bibnamefont
  {Kiehn}}, \bibinfo {author} {\bibfnamefont {V.~P.}\ \bibnamefont {Singh}},\
  and\ \bibinfo {author} {\bibfnamefont {L.}~\bibnamefont {Mathey}},\
  }\bibfield  {title} {\bibinfo {title} {Implementation of an atomtronic squid
  in a strongly confined toroidal condensate},\ }\href
  {https://doi.org/10.1103/PhysRevResearch.4.033024} {\bibfield  {journal}
  {\bibinfo  {journal} {Phys. Rev. Res.}\ }\textbf {\bibinfo {volume} {4}},\
  \bibinfo {pages} {033024} (\bibinfo {year} {2022})}\BibitemShut {NoStop}%
\bibitem [{\citenamefont {Arivazhagan}\ \emph {et~al.}(2023)\citenamefont
  {Arivazhagan}, \citenamefont {Muruganandam},\ and\ \citenamefont
  {Athavan}}]{ARIVAZHAGAN20231354180}%
  \BibitemOpen
  \bibfield  {author} {\bibinfo {author} {\bibfnamefont {M.}~\bibnamefont
  {Arivazhagan}}, \bibinfo {author} {\bibfnamefont {P.}~\bibnamefont
  {Muruganandam}},\ and\ \bibinfo {author} {\bibfnamefont {N.}~\bibnamefont
  {Athavan}},\ }\bibfield  {title} {\bibinfo {title} {Parametric triggering of
  vortices in toroidally trapped rotating bose–einstein condensates},\ }\href
  {https://doi.org/https://doi.org/10.1016/j.physc.2022.1354180} {\bibfield
  {journal} {\bibinfo  {journal} {Physica C: Superconductivity and its
  Applications}\ }\textbf {\bibinfo {volume} {604}},\ \bibinfo {pages}
  {1354180} (\bibinfo {year} {2023})}\BibitemShut {NoStop}%
\bibitem [{\citenamefont {Pradhan}\ \emph {et~al.}(2024)\citenamefont
  {Pradhan}, \citenamefont {Kumar}, \citenamefont {Kanamoto}, \citenamefont
  {Dey}, \citenamefont {Bhattacharya},\ and\ \citenamefont
  {Mishra}}]{PhysRevResearch.6.013104}%
  \BibitemOpen
  \bibfield  {author} {\bibinfo {author} {\bibfnamefont {N.}~\bibnamefont
  {Pradhan}}, \bibinfo {author} {\bibfnamefont {P.}~\bibnamefont {Kumar}},
  \bibinfo {author} {\bibfnamefont {R.}~\bibnamefont {Kanamoto}}, \bibinfo
  {author} {\bibfnamefont {T.~N.}\ \bibnamefont {Dey}}, \bibinfo {author}
  {\bibfnamefont {M.}~\bibnamefont {Bhattacharya}},\ and\ \bibinfo {author}
  {\bibfnamefont {P.~K.}\ \bibnamefont {Mishra}},\ }\bibfield  {title}
  {\bibinfo {title} {Cavity optomechanical detection of persistent currents and
  solitons in a bosonic ring condensate},\ }\href
  {https://doi.org/10.1103/PhysRevResearch.6.013104} {\bibfield  {journal}
  {\bibinfo  {journal} {Phys. Rev. Res.}\ }\textbf {\bibinfo {volume} {6}},\
  \bibinfo {pages} {013104} (\bibinfo {year} {2024})}\BibitemShut {NoStop}%
\bibitem [{\citenamefont {Xhani}\ \emph {et~al.}(2023)\citenamefont {Xhani},
  \citenamefont {Del~Pace}, \citenamefont {Scazza},\ and\ \citenamefont
  {Roati}}]{atoms11080109}%
  \BibitemOpen
  \bibfield  {author} {\bibinfo {author} {\bibfnamefont {K.}~\bibnamefont
  {Xhani}}, \bibinfo {author} {\bibfnamefont {G.}~\bibnamefont {Del~Pace}},
  \bibinfo {author} {\bibfnamefont {F.}~\bibnamefont {Scazza}},\ and\ \bibinfo
  {author} {\bibfnamefont {G.}~\bibnamefont {Roati}},\ }\bibfield  {title}
  {\bibinfo {title} {Decay of persistent currents in annular atomic
  superfluids},\ }\bibfield  {journal} {\bibinfo  {journal} {Atoms}\ }\textbf
  {\bibinfo {volume} {11}},\ \href {https://doi.org/10.3390/atoms11080109}
  {10.3390/atoms11080109} (\bibinfo {year} {2023})\BibitemShut {NoStop}%
\bibitem [{\citenamefont {Pecci}\ \emph {et~al.}(2023)\citenamefont {Pecci},
  \citenamefont {Aupetit-Diallo}, \citenamefont {Albert}, \citenamefont
  {Vignolo},\ and\ \citenamefont {Minguzzi}}]{CRPHYS_2023__24_S3_A6_0}%
  \BibitemOpen
  \bibfield  {author} {\bibinfo {author} {\bibfnamefont {G.}~\bibnamefont
  {Pecci}}, \bibinfo {author} {\bibfnamefont {G.}~\bibnamefont
  {Aupetit-Diallo}}, \bibinfo {author} {\bibfnamefont {M.}~\bibnamefont
  {Albert}}, \bibinfo {author} {\bibfnamefont {P.}~\bibnamefont {Vignolo}},\
  and\ \bibinfo {author} {\bibfnamefont {A.}~\bibnamefont {Minguzzi}},\
  }\bibfield  {title} {\bibinfo {title} {Persistent currents in a strongly
  interacting multicomponent {Bose} gas on a ring},\ }\bibfield  {journal}
  {\bibinfo  {journal} {Comptes Rendus. Physique}\ }\href
  {https://doi.org/10.5802/crphys.157} {10.5802/crphys.157} (\bibinfo {year}
  {2023}),\ \bibinfo {note} {online first}\BibitemShut {NoStop}%
\bibitem [{\citenamefont {Tekverk}\ \emph {et~al.}(2024)\citenamefont
  {Tekverk}, \citenamefont {Siebor},\ and\ \citenamefont
  {Das}}]{PhysRevA.109.023315}%
  \BibitemOpen
  \bibfield  {author} {\bibinfo {author} {\bibfnamefont {J.}~\bibnamefont
  {Tekverk}}, \bibinfo {author} {\bibfnamefont {C.}~\bibnamefont {Siebor}},\
  and\ \bibinfo {author} {\bibfnamefont {K.~K.}\ \bibnamefont {Das}},\
  }\bibfield  {title} {\bibinfo {title} {Effects of a rotating periodic lattice
  on coherent quantum states in a ring topology: The case of negative
  nonlinearity},\ }\href {https://doi.org/10.1103/PhysRevA.109.023315}
  {\bibfield  {journal} {\bibinfo  {journal} {Phys. Rev. A}\ }\textbf {\bibinfo
  {volume} {109}},\ \bibinfo {pages} {023315} (\bibinfo {year}
  {2024})}\BibitemShut {NoStop}%
\bibitem [{\citenamefont {Mathey}\ \emph {et~al.}(2014)\citenamefont {Mathey},
  \citenamefont {Clark},\ and\ \citenamefont {Mathey}}]{TWA}%
  \BibitemOpen
  \bibfield  {author} {\bibinfo {author} {\bibfnamefont {A.~C.}\ \bibnamefont
  {Mathey}}, \bibinfo {author} {\bibfnamefont {C.~W.}\ \bibnamefont {Clark}},\
  and\ \bibinfo {author} {\bibfnamefont {L.}~\bibnamefont {Mathey}},\
  }\bibfield  {title} {\bibinfo {title} {Decay of a superfluid current of
  ultracold atoms in a toroidal trap},\ }\href
  {https://doi.org/10.1103/PhysRevA.90.023604} {\bibfield  {journal} {\bibinfo
  {journal} {Phys. Rev. A}\ }\textbf {\bibinfo {volume} {90}},\ \bibinfo
  {pages} {023604} (\bibinfo {year} {2014})}\BibitemShut {NoStop}%
\bibitem [{\citenamefont {Fisher}\ \emph {et~al.}(1989)\citenamefont {Fisher},
  \citenamefont {Weichman}, \citenamefont {Grinstein},\ and\ \citenamefont
  {Fisher}}]{Fisher1989}%
  \BibitemOpen
  \bibfield  {author} {\bibinfo {author} {\bibfnamefont {M.~P.~A.}\
  \bibnamefont {Fisher}}, \bibinfo {author} {\bibfnamefont {P.~B.}\
  \bibnamefont {Weichman}}, \bibinfo {author} {\bibfnamefont {G.}~\bibnamefont
  {Grinstein}},\ and\ \bibinfo {author} {\bibfnamefont {D.~S.}\ \bibnamefont
  {Fisher}},\ }\bibfield  {title} {\bibinfo {title} {Boson localization and the
  superfluid-insulator transition},\ }\href
  {https://doi.org/10.1103/PhysRevB.40.546} {\bibfield  {journal} {\bibinfo
  {journal} {Phys. Rev. B}\ }\textbf {\bibinfo {volume} {40}},\ \bibinfo
  {pages} {546} (\bibinfo {year} {1989})}\BibitemShut {NoStop}%
\bibitem [{\citenamefont {Freericks}\ and\ \citenamefont
  {Monien}(1994)}]{Freericks1994}%
  \BibitemOpen
  \bibfield  {author} {\bibinfo {author} {\bibfnamefont {J.~K.}\ \bibnamefont
  {Freericks}}\ and\ \bibinfo {author} {\bibfnamefont {H.}~\bibnamefont
  {Monien}},\ }\bibfield  {title} {\bibinfo {title} {Phase diagram of the
  bose-hubbard model},\ }\href {https://doi.org/10.1209/0295-5075/26/7/012}
  {\bibfield  {journal} {\bibinfo  {journal} {Europhysics Letters}\ }\textbf
  {\bibinfo {volume} {26}},\ \bibinfo {pages} {545} (\bibinfo {year}
  {1994})}\BibitemShut {NoStop}%
\bibitem [{\citenamefont {Freericks}\ \emph {et~al.}(2009)\citenamefont
  {Freericks}, \citenamefont {Krishnamurthy}, \citenamefont {Kato},
  \citenamefont {Kawashima},\ and\ \citenamefont {Trivedi}}]{HRK2009}%
  \BibitemOpen
  \bibfield  {author} {\bibinfo {author} {\bibfnamefont {J.~K.}\ \bibnamefont
  {Freericks}}, \bibinfo {author} {\bibfnamefont {H.~R.}\ \bibnamefont
  {Krishnamurthy}}, \bibinfo {author} {\bibfnamefont {Y.}~\bibnamefont {Kato}},
  \bibinfo {author} {\bibfnamefont {N.}~\bibnamefont {Kawashima}},\ and\
  \bibinfo {author} {\bibfnamefont {N.}~\bibnamefont {Trivedi}},\ }\bibfield
  {title} {\bibinfo {title} {Strong-coupling expansion for the momentum
  distribution of the bose-hubbard model with benchmarking against exact
  numerical results},\ }\href {https://doi.org/10.1103/PhysRevA.79.053631}
  {\bibfield  {journal} {\bibinfo  {journal} {Phys. Rev. A}\ }\textbf {\bibinfo
  {volume} {79}},\ \bibinfo {pages} {053631} (\bibinfo {year}
  {2009})}\BibitemShut {NoStop}%
\bibitem [{\citenamefont {Gupta}\ \emph {et~al.}(2013)\citenamefont {Gupta},
  \citenamefont {Krishnamurthy},\ and\ \citenamefont {Freericks}}]{Manjari}%
  \BibitemOpen
  \bibfield  {author} {\bibinfo {author} {\bibfnamefont {M.}~\bibnamefont
  {Gupta}}, \bibinfo {author} {\bibfnamefont {H.~R.}\ \bibnamefont
  {Krishnamurthy}},\ and\ \bibinfo {author} {\bibfnamefont {J.~K.}\
  \bibnamefont {Freericks}},\ }\bibfield  {title} {\bibinfo {title}
  {Strong-coupling expansion for ultracold bosons in an optical lattice at
  finite temperatures in the presence of superfluidity},\ }\href
  {https://doi.org/10.1103/PhysRevA.88.053636} {\bibfield  {journal} {\bibinfo
  {journal} {Phys. Rev. A}\ }\textbf {\bibinfo {volume} {88}},\ \bibinfo
  {pages} {053636} (\bibinfo {year} {2013})}\BibitemShut {NoStop}%
\bibitem [{\citenamefont {Aghamalyan}\ \emph
  {et~al.}(2015{\natexlab{a}})\citenamefont {Aghamalyan}, \citenamefont
  {Cominotti}, \citenamefont {Rizzi}, \citenamefont {Rossini}, \citenamefont
  {Hekking}, \citenamefont {Minguzzi}, \citenamefont {Kwek},\ and\
  \citenamefont {Amico}}]{SqubitT}%
  \BibitemOpen
  \bibfield  {author} {\bibinfo {author} {\bibfnamefont {D.}~\bibnamefont
  {Aghamalyan}}, \bibinfo {author} {\bibfnamefont {M.}~\bibnamefont
  {Cominotti}}, \bibinfo {author} {\bibfnamefont {M.}~\bibnamefont {Rizzi}},
  \bibinfo {author} {\bibfnamefont {D.}~\bibnamefont {Rossini}}, \bibinfo
  {author} {\bibfnamefont {F.}~\bibnamefont {Hekking}}, \bibinfo {author}
  {\bibfnamefont {A.}~\bibnamefont {Minguzzi}}, \bibinfo {author}
  {\bibfnamefont {L.-C.}\ \bibnamefont {Kwek}},\ and\ \bibinfo {author}
  {\bibfnamefont {L.}~\bibnamefont {Amico}},\ }\bibfield  {title} {\bibinfo
  {title} {Coherent superposition of current flows in an atomtronic quantum
  interference device},\ }\href@noop {} {\bibfield  {journal} {\bibinfo
  {journal} {New Journal of Physics}\ }\textbf {\bibinfo {volume} {17}},\
  \bibinfo {pages} {045023} (\bibinfo {year} {2015}{\natexlab{a}})}\BibitemShut
  {NoStop}%
\bibitem [{\citenamefont {Hettiarachchilage}\ \emph {et~al.}(2013)\citenamefont
  {Hettiarachchilage}, \citenamefont {Rousseau}, \citenamefont {Tam},
  \citenamefont {Jarrell},\ and\ \citenamefont {Moreno}}]{PhasDiag}%
  \BibitemOpen
  \bibfield  {author} {\bibinfo {author} {\bibfnamefont {K.}~\bibnamefont
  {Hettiarachchilage}}, \bibinfo {author} {\bibfnamefont {V.~G.}\ \bibnamefont
  {Rousseau}}, \bibinfo {author} {\bibfnamefont {K.-M.}\ \bibnamefont {Tam}},
  \bibinfo {author} {\bibfnamefont {M.}~\bibnamefont {Jarrell}},\ and\ \bibinfo
  {author} {\bibfnamefont {J.}~\bibnamefont {Moreno}},\ }\bibfield  {title}
  {\bibinfo {title} {Phase diagram of the bose-hubbard model on a ring-shaped
  lattice with tunable weak links},\ }\href
  {https://doi.org/10.1103/PhysRevA.87.051607} {\bibfield  {journal} {\bibinfo
  {journal} {Phys. Rev. A}\ }\textbf {\bibinfo {volume} {87}},\ \bibinfo
  {pages} {051607} (\bibinfo {year} {2013})}\BibitemShut {NoStop}%
\bibitem [{\citenamefont {Arwas}\ and\ \citenamefont
  {Cohen}(2017)}]{PhysRevB.95.054505}%
  \BibitemOpen
  \bibfield  {author} {\bibinfo {author} {\bibfnamefont {G.}~\bibnamefont
  {Arwas}}\ and\ \bibinfo {author} {\bibfnamefont {D.}~\bibnamefont {Cohen}},\
  }\bibfield  {title} {\bibinfo {title} {Superfluidity in bose-hubbard
  circuits},\ }\href {https://doi.org/10.1103/PhysRevB.95.054505} {\bibfield
  {journal} {\bibinfo  {journal} {Phys. Rev. B}\ }\textbf {\bibinfo {volume}
  {95}},\ \bibinfo {pages} {054505} (\bibinfo {year} {2017})}\BibitemShut
  {NoStop}%
\bibitem [{\citenamefont {Kohn}\ \emph {et~al.}(2020)\citenamefont {Kohn},
  \citenamefont {Silvi}, \citenamefont {Gerster}, \citenamefont {Keck},
  \citenamefont {Fazio}, \citenamefont {Santoro},\ and\ \citenamefont
  {Montangero}}]{PhysRevA.101.023617}%
  \BibitemOpen
  \bibfield  {author} {\bibinfo {author} {\bibfnamefont {L.}~\bibnamefont
  {Kohn}}, \bibinfo {author} {\bibfnamefont {P.}~\bibnamefont {Silvi}},
  \bibinfo {author} {\bibfnamefont {M.}~\bibnamefont {Gerster}}, \bibinfo
  {author} {\bibfnamefont {M.}~\bibnamefont {Keck}}, \bibinfo {author}
  {\bibfnamefont {R.}~\bibnamefont {Fazio}}, \bibinfo {author} {\bibfnamefont
  {G.~E.}\ \bibnamefont {Santoro}},\ and\ \bibinfo {author} {\bibfnamefont
  {S.}~\bibnamefont {Montangero}},\ }\bibfield  {title} {\bibinfo {title}
  {Superfluid-to-mott transition in a bose-hubbard ring: Persistent currents
  and defect formation},\ }\href {https://doi.org/10.1103/PhysRevA.101.023617}
  {\bibfield  {journal} {\bibinfo  {journal} {Phys. Rev. A}\ }\textbf {\bibinfo
  {volume} {101}},\ \bibinfo {pages} {023617} (\bibinfo {year}
  {2020})}\BibitemShut {NoStop}%
\bibitem [{\citenamefont {Tengstrand}\ \emph {et~al.}(2021)\citenamefont
  {Tengstrand}, \citenamefont {Boholm}, \citenamefont {Sachdeva}, \citenamefont
  {Bengtsson},\ and\ \citenamefont {Reimann}}]{PhysRevA.103.013313}%
  \BibitemOpen
  \bibfield  {author} {\bibinfo {author} {\bibfnamefont {M.~N.}\ \bibnamefont
  {Tengstrand}}, \bibinfo {author} {\bibfnamefont {D.}~\bibnamefont {Boholm}},
  \bibinfo {author} {\bibfnamefont {R.}~\bibnamefont {Sachdeva}}, \bibinfo
  {author} {\bibfnamefont {J.}~\bibnamefont {Bengtsson}},\ and\ \bibinfo
  {author} {\bibfnamefont {S.~M.}\ \bibnamefont {Reimann}},\ }\bibfield
  {title} {\bibinfo {title} {Persistent currents in toroidal dipolar
  supersolids},\ }\href {https://doi.org/10.1103/PhysRevA.103.013313}
  {\bibfield  {journal} {\bibinfo  {journal} {Phys. Rev. A}\ }\textbf {\bibinfo
  {volume} {103}},\ \bibinfo {pages} {013313} (\bibinfo {year}
  {2021})}\BibitemShut {NoStop}%
\bibitem [{\citenamefont {P\^a\ifmmode~\mbox{\c{t}}\else \c{t}\fi{}u}\ and\
  \citenamefont {Averin}(2022)}]{PhysRevLett.128.096801}%
  \BibitemOpen
  \bibfield  {author} {\bibinfo {author} {\bibfnamefont {O.~I.}\ \bibnamefont
  {P\^a\ifmmode~\mbox{\c{t}}\else \c{t}\fi{}u}}\ and\ \bibinfo {author}
  {\bibfnamefont {D.~V.}\ \bibnamefont {Averin}},\ }\bibfield  {title}
  {\bibinfo {title} {Temperature-dependent periodicity of the persistent
  current in strongly interacting systems},\ }\href
  {https://doi.org/10.1103/PhysRevLett.128.096801} {\bibfield  {journal}
  {\bibinfo  {journal} {Phys. Rev. Lett.}\ }\textbf {\bibinfo {volume} {128}},\
  \bibinfo {pages} {096801} (\bibinfo {year} {2022})}\BibitemShut {NoStop}%
\bibitem [{\citenamefont {Moulder}\ \emph {et~al.}(2012)\citenamefont
  {Moulder}, \citenamefont {Beattie}, \citenamefont {Smith}, \citenamefont
  {Tammuz},\ and\ \citenamefont {Hadzibabic}}]{QSDnDecay}%
  \BibitemOpen
  \bibfield  {author} {\bibinfo {author} {\bibfnamefont {S.}~\bibnamefont
  {Moulder}}, \bibinfo {author} {\bibfnamefont {S.}~\bibnamefont {Beattie}},
  \bibinfo {author} {\bibfnamefont {R.~P.}\ \bibnamefont {Smith}}, \bibinfo
  {author} {\bibfnamefont {N.}~\bibnamefont {Tammuz}},\ and\ \bibinfo {author}
  {\bibfnamefont {Z.}~\bibnamefont {Hadzibabic}},\ }\bibfield  {title}
  {\bibinfo {title} {Quantized supercurrent decay in an annular bose-einstein
  condensate},\ }\href {https://doi.org/10.1103/PhysRevA.86.013629} {\bibfield
  {journal} {\bibinfo  {journal} {Phys. Rev. A}\ }\textbf {\bibinfo {volume}
  {86}},\ \bibinfo {pages} {013629} (\bibinfo {year} {2012})}\BibitemShut
  {NoStop}%
\bibitem [{\citenamefont {Yakimenko}\ \emph {et~al.}(2015)\citenamefont
  {Yakimenko}, \citenamefont {Bidasyuk}, \citenamefont {Weyrauch},
  \citenamefont {Kuriatnikov},\ and\ \citenamefont
  {Vilchinskii}}]{PhysRevA.91.033607}%
  \BibitemOpen
  \bibfield  {author} {\bibinfo {author} {\bibfnamefont {A.~I.}\ \bibnamefont
  {Yakimenko}}, \bibinfo {author} {\bibfnamefont {Y.~M.}\ \bibnamefont
  {Bidasyuk}}, \bibinfo {author} {\bibfnamefont {M.}~\bibnamefont {Weyrauch}},
  \bibinfo {author} {\bibfnamefont {Y.~I.}\ \bibnamefont {Kuriatnikov}},\ and\
  \bibinfo {author} {\bibfnamefont {S.~I.}\ \bibnamefont {Vilchinskii}},\
  }\bibfield  {title} {\bibinfo {title} {Vortices in a toroidal bose-einstein
  condensate with a rotating weak link},\ }\href
  {https://doi.org/10.1103/PhysRevA.91.033607} {\bibfield  {journal} {\bibinfo
  {journal} {Phys. Rev. A}\ }\textbf {\bibinfo {volume} {91}},\ \bibinfo
  {pages} {033607} (\bibinfo {year} {2015})}\BibitemShut {NoStop}%
\bibitem [{\citenamefont {Kunimi}\ and\ \citenamefont
  {Danshita}(2019)}]{PhysRevA.99.043613}%
  \BibitemOpen
  \bibfield  {author} {\bibinfo {author} {\bibfnamefont {M.}~\bibnamefont
  {Kunimi}}\ and\ \bibinfo {author} {\bibfnamefont {I.}~\bibnamefont
  {Danshita}},\ }\bibfield  {title} {\bibinfo {title} {Decay mechanisms of
  superflow of bose-einstein condensates in ring traps},\ }\href
  {https://doi.org/10.1103/PhysRevA.99.043613} {\bibfield  {journal} {\bibinfo
  {journal} {Phys. Rev. A}\ }\textbf {\bibinfo {volume} {99}},\ \bibinfo
  {pages} {043613} (\bibinfo {year} {2019})}\BibitemShut {NoStop}%
\bibitem [{\citenamefont {Mehdi}\ \emph {et~al.}(2021)\citenamefont {Mehdi},
  \citenamefont {Bradley}, \citenamefont {Hope},\ and\ \citenamefont
  {Szigeti}}]{10.21468/SciPostPhys.11.4.080}%
  \BibitemOpen
  \bibfield  {author} {\bibinfo {author} {\bibfnamefont {Z.}~\bibnamefont
  {Mehdi}}, \bibinfo {author} {\bibfnamefont {A.~S.}\ \bibnamefont {Bradley}},
  \bibinfo {author} {\bibfnamefont {J.~J.}\ \bibnamefont {Hope}},\ and\
  \bibinfo {author} {\bibfnamefont {S.~S.}\ \bibnamefont {Szigeti}},\
  }\bibfield  {title} {\bibinfo {title} {Superflow decay in a toroidal bose
  gas: The effect of quantum and thermal fluctuations},\ }\href
  {https://doi.org/10.21468/SciPostPhys.11.4.080} {\bibfield  {journal}
  {\bibinfo  {journal} {SciPost Phys.}\ }\textbf {\bibinfo {volume} {11}},\
  \bibinfo {pages} {080} (\bibinfo {year} {2021})}\BibitemShut {NoStop}%
\bibitem [{\citenamefont {Aghamalyan}\ \emph
  {et~al.}(2015{\natexlab{b}})\citenamefont {Aghamalyan}, \citenamefont
  {Cominotti}, \citenamefont {Rizzi}, \citenamefont {Rossini}, \citenamefont
  {Hekking}, \citenamefont {Minguzzi}, \citenamefont {Kwek},\ and\
  \citenamefont {Amico}}]{Aghamalyan_2015}%
  \BibitemOpen
  \bibfield  {author} {\bibinfo {author} {\bibfnamefont {D.}~\bibnamefont
  {Aghamalyan}}, \bibinfo {author} {\bibfnamefont {M.}~\bibnamefont
  {Cominotti}}, \bibinfo {author} {\bibfnamefont {M.}~\bibnamefont {Rizzi}},
  \bibinfo {author} {\bibfnamefont {D.}~\bibnamefont {Rossini}}, \bibinfo
  {author} {\bibfnamefont {F.}~\bibnamefont {Hekking}}, \bibinfo {author}
  {\bibfnamefont {A.}~\bibnamefont {Minguzzi}}, \bibinfo {author}
  {\bibfnamefont {L.-C.}\ \bibnamefont {Kwek}},\ and\ \bibinfo {author}
  {\bibfnamefont {L.}~\bibnamefont {Amico}},\ }\bibfield  {title} {\bibinfo
  {title} {Coherent superposition of current flows in an atomtronic quantum
  interference device},\ }\href {https://doi.org/10.1088/1367-2630/17/4/045023}
  {\bibfield  {journal} {\bibinfo  {journal} {New Journal of Physics}\ }\textbf
  {\bibinfo {volume} {17}},\ \bibinfo {pages} {045023} (\bibinfo {year}
  {2015}{\natexlab{b}})}\BibitemShut {NoStop}%
\bibitem [{\citenamefont {Aghamalyan}\ \emph {et~al.}(2016)\citenamefont
  {Aghamalyan}, \citenamefont {Nguyen}, \citenamefont {Auksztol}, \citenamefont
  {Gan}, \citenamefont {Valado}, \citenamefont {Condylis}, \citenamefont
  {Kwek}, \citenamefont {Dumke},\ and\ \citenamefont
  {Amico}}]{Aghamalyan_2016}%
  \BibitemOpen
  \bibfield  {author} {\bibinfo {author} {\bibfnamefont {D.}~\bibnamefont
  {Aghamalyan}}, \bibinfo {author} {\bibfnamefont {N.~T.}\ \bibnamefont
  {Nguyen}}, \bibinfo {author} {\bibfnamefont {F.}~\bibnamefont {Auksztol}},
  \bibinfo {author} {\bibfnamefont {K.~S.}\ \bibnamefont {Gan}}, \bibinfo
  {author} {\bibfnamefont {M.~M.}\ \bibnamefont {Valado}}, \bibinfo {author}
  {\bibfnamefont {P.~C.}\ \bibnamefont {Condylis}}, \bibinfo {author}
  {\bibfnamefont {L.-C.}\ \bibnamefont {Kwek}}, \bibinfo {author}
  {\bibfnamefont {R.}~\bibnamefont {Dumke}},\ and\ \bibinfo {author}
  {\bibfnamefont {L.}~\bibnamefont {Amico}},\ }\bibfield  {title} {\bibinfo
  {title} {An atomtronic flux qubit: a ring lattice of bose–einstein
  condensates interrupted by three weak links},\ }\href
  {https://doi.org/10.1088/1367-2630/18/7/075013} {\bibfield  {journal}
  {\bibinfo  {journal} {New Journal of Physics}\ }\textbf {\bibinfo {volume}
  {18}},\ \bibinfo {pages} {075013} (\bibinfo {year} {2016})}\BibitemShut
  {NoStop}%
\end{thebibliography}%

\end{document}